\documentclass[aps,prb,amsmath,amssymb,showpacs,twocolumn,superscriptaddress]{revtex4-2}
\usepackage{amsmath}
\usepackage{amssymb}
\usepackage{amsthm}
\usepackage{setspace}
\usepackage{graphicx}
\usepackage{braket}
\usepackage{mathrsfs}
\usepackage{float}
\usepackage{natbib}
\usepackage[colorlinks = true,linkcolor = blue,urlcolor  = blue,citecolor = blue,anchorcolor = blue]{hyperref}
\usepackage[utf8]{inputenc}
\usepackage[english]{babel}
\usepackage{bm}
\usepackage{csquotes}

\begin{document}

\title{Chiral anomaly-induced nonlinear Hall effect in spin-orbit coupled noncentrosymmetric metals}
\author{Gautham Varma K.}
\affiliation{School of Physical Sciences, Indian Institute of Technology Mandi, Mandi-175005, India.}
\author{Mohd.~Hashim Raza}
\affiliation{Maharani Lal Kunwari College, Tulsipur Road, Balrampur, Uttar pradesh-271201, India.}
\author{Azaz Ahmad}
\affiliation{School of Physical Sciences, Indian Institute of Technology Mandi, Mandi-175005, India.}
\begin{abstract}
Recent studies have shown that chiral anomaly is not limited to WSMs, but are also shown by a larger class of materials called spin orbit coupled noncentrosymmetric metals (SOC-NCMs), which has shed more insight into the origin of chiral anomaly as a Fermi surface property rather than a nodal property. In this study, we explore nonlinear transport responses in SOC-NCMs within the framework of semiclassical dynamics, employing the Maxwell-Boltzmann transport theory augmented by charge conservation and momentum-dependent scattering processes. We take into account both non-magnetic and magnetic impurity scattering mechanisms. We demonstrate that the chiral-anomaly-induced nonlinear Hall (CNLH) response exhibits a characteristic quadratic dependence on the applied magnetic field and remains negative for both types of impurities. We find that magnetic scatterers leading to enhanced/suppressed interband scattering modifies the magnitude of the signal, but does not affect its qualitative behavior. In contrast, the presence of tilt in the band dispersion induces a pronounced anisotropic response, including a magnetic-field-direction-dependent sign reversal that can be categorized into ‘weak and strong’ regimes. Furthermore, the CNLH response shows substantial directional anisotropy governed by the relative orientation of the external magnetic field and the tilt vector. Our findings will be helpful in designing the experimental setup to get direction-dependent conductivity, which can be tuned externally with the help of magnetic impurity sites.

\end{abstract}
 
\maketitle
\section{Introduction} \label{I_Introduction}

Weyl fermions, ever since it has been discovered in the context of high energy physics, took a long time to be realized in condensed matter systems as quasiparticles~\cite{bevan1997momentum,wan2011topological,Yan_2017,armitage2018weyl,burkov2011weyl}. Its realization as collective excitations in condensed matter systems, provides a realistic platform to test the prediction of the Weyl equation originally introduced in the high energy context~\cite{1929ZPhy...56..330W,abers2004quantum}. These materials are dubbed Weyl semimetals and are a very focused area of investigation among researchers due to their fascinating properties that were generally absent in other materials. Some of the very celebrated properties include open Fermi arcs ~\cite{wan2011topological}, anomalous Hall \cite{yang2011quantum,burkov2014anomalous} and Nernst effects \cite{sharma2016nernst,sharma2017nernst,liang2017anomalous}, planar Hall and Nernst effects \cite{nandy2017chiral,sharma2019transverse}, and most important one is the manifestation of chiral or Adler-Bell-Jackiw anomaly \cite{adler1969axial,nielsen1981no,nielsen1983adler,bell1969pcac,aji2012adler,zyuzin2012weyl,zyuzin2012weyl,son2012berry,goswami2015optical, fukushima2008chiral,goswami2013axionic}. Weyl fermions have a very intrinsic property associated with them, i.e., chirality $\chi$ ($=\pm1$), which can be calculated by performing a closed surface integral of Berry flux around the band touching points (Weyl point). The number of Weyl Fermions associated with a particular chirality is fixed, leading to the electronic as well as chiral charge conservation. However, this does not hold when the system is brought under external gauge fields such as electric and magnetic fields. This is called chiral anomaly(CA).\\
Weyl semimetals such as TaAs show negative magnetoresistance (positive magnetoconductance) induced by chiral anomaly \cite{son2013chiral,kim2014boltzmann,burkov2014anomalous} that has been verified experimentally \cite{zhang2016linear,fukushima2008chiral,zyuzin2012topological}. In the absence of an electric field, an equivalent number of Weyl fermions exhibiting opposite chiralities move in opposite directions (aligned with the external magnetic field). This configuration leads to a null net charge current. However, when an electric field is applied parallel to the magnetic field, a discernible chemical potential difference arises between Weyl fermions with opposite chiralities, a consequence of the chiral anomaly \cite{son2013chiral}. This results in an imbalance between the two streams of Weyl fermions and, consequently, gives rise to a net charge current represented by $\mathbf{j}\propto (\mathbf{E}\cdot\mathbf{B)}\mathbf{B}$.\\

Recently, CA was predicted to occur in materials where only one effective Weyl node is present with two disjointed and concentric Fermi surfaces having opposing Berry flux associated with it~\cite{das2023chiral,bradlyn2016beyond,chang2018topological,hasan2021weyl}. These Weyl nodes are dubbed as `Kramers-Weyl nodes (KWNs) and are found at time-reversal-invariant momentum (TRIM) points \cite{bradlyn2016beyond,chang2018topological,hasan2021weyl}. The materials that hosts such nodes are called spin orbit coupled noncentrosymmetric metals (SOC-NCMs). Like WSMs, they too have a $\mathbf{k}\cdot\sigma$ term in the Hamiltonian but with an additional $\mathbf{k}^2$ term. However, unlike WSMs, the $\mathbf{\sigma}$ matrices here represents real spins. The Chern number turns out to be non-zero corresponding to the two Fermi surfaces, leading to a total sum still equal to zero \cite{cheon2022chiral,gao2022chiral}. The recent findings advance the understanding of the CA induced linear as well as non-linear transport properties of SOC-NCMs~\cite{varma2024magnetotransport,ahmad2025chiral,verma2019thermoelectric,sarkar2025symmetrydrivenintrinsicnonlinearpure,mandal2025disentangling,ahmad2025longitudinal}. Although CA is demonstrated by both SOC-NCMs and WSMs, their transport responses are remarkably distinct \cite{varma2024magnetotransport,ahmad2025chiral}.
The non-linear Hall effect in WSM induced by chiral anomaly has been introduced by \textit{Rui-Hao Li et al.} in Ref.~\cite{PhysRevB.103.045105}. Authors have found that the inversion symmetry-broken WSM having a tilt of the Weyl cone will have a nonlinear Hall effect induced by the chiral anomaly and dubbed it as `Chiral anomaly induced nonlinear Hall effect (CNLH)'. This CNLH is the combined effect of anomalous velocity $v^{\chi}_{a} = - \frac{e}{\hbar}\mathbf{E}\times \mathbf{\Omega^{\chi}_{k}}$ and chiral anomaly \cite{PhysRev.95.1154,xiao2010berry}. To get CNLH, one has to integrate anomalous velocity on the Fermi surface~\cite{PhysRevB.103.045105,ahmad2025chiral}. The CNLH effect is nonzero provided that the Fermi surface is asymmetric and the Hamiltonian has broken inversion symmetry. The tilt of the Weyl cone provides an asymmetric Fermi surface about the projection of the Weyl node on the Fermi surface~\cite{PhysRevB.103.045105}. Here, it is crucial to note that CNLH is distinct from other non-linear Hall effects caused by the Berry curvature dipole (BCD) \cite{PhysRevB.97.201404,doi:10.1126/sciadv.1603266} or quadrupole \cite{zhang2024third}, as the latter can persist even without an external magnetic field.

\begin{figure}
    \includegraphics[width=0.8\columnwidth]{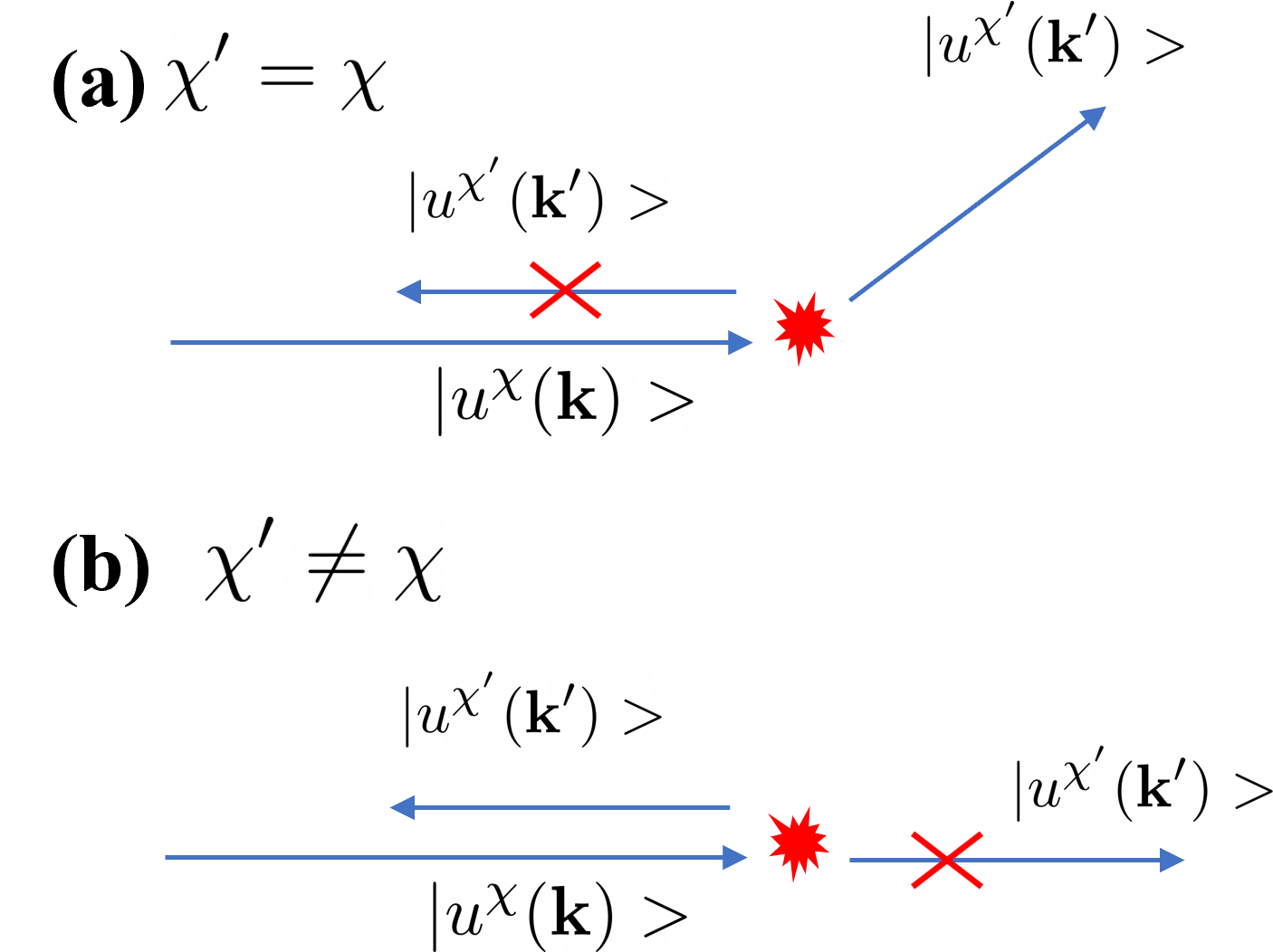}
    \caption{Schematic representation of the back-scattering process in the presence of non-magnetic impurity potential represented by the red star. Back-scattering is intrinsically suppressed for chirality-conserving scattering events due to the topological protection of Weyl fermions, whereas it becomes allowed in chirality-flipping processes.}
\label{Fig:Backscattering_Schematic}
\end{figure}

In previous works related to CNLH ~\cite{li2021nonlinear, nandy2021chiral,zeng2022chiral}, it is inherently assumed that the interband scattering rate is much lower than the intraband scattering rate. However, recent studies show that it is crucial to include interband scattering to naturally explain chiral anomaly induced charge transport~\cite{sharma2023decoupling,ahmad2021longitudinal,ahmad2023longitudinal,sharma2016nernst,sharma2017chiral,sharma2017nernst,sharma2019transverse,sharma2020sign,ahmad2025chiral,varma2024magnetotransport,ahmad2024geometry}. Motivated by the impact of CA on transport properties in WSMs, under semiclassical treatment with chiral charge as well as number conservation, we have done a comprehensive analysis of CNLH in SOC-NCMs by going beyond the constant relaxation time approximation. Additionally, we assume that the system is stationary and uniform so that the distribution function will merely be momentum--dependent. In the present study, for the sake of completeness as well as for experimental usefulness, we have included magnetic as well as non-magnetic impurity scattering sites. The presence of magnetic impurity sites, which are magnetically active, can flip the spin of the incoming wave packet. This plays a very important role in the CA-assisted transport. In Fig.~\ref{Fig:Backscattering_Schematic}, we schematically draw the scattering mechanism for both cases: (i) chirality preserving (intraband) and (ii) chirality breaking (interband) scattering. Overlap of the wavefunction between incoming and outgoing wavefunctions has a different functional form: non-magnetic impurities ($\sigma_0$) preserve the spin structure, while magnetic impurities ($\sigma_i,i=x,y,z$) cause spin-dependent scattering. This leads to interesting transport properties that provide a deeper understanding of the CNLH effect, that captures the missing facts in the previous studies, and will be more beneficial in elucidating the nonlinear conductivities that can appear via nonlinear transport measurement and nonlinear optical phenomena such as photocurrent and second harmonic production.
\begin{figure}
    \includegraphics[width=0.95\columnwidth]{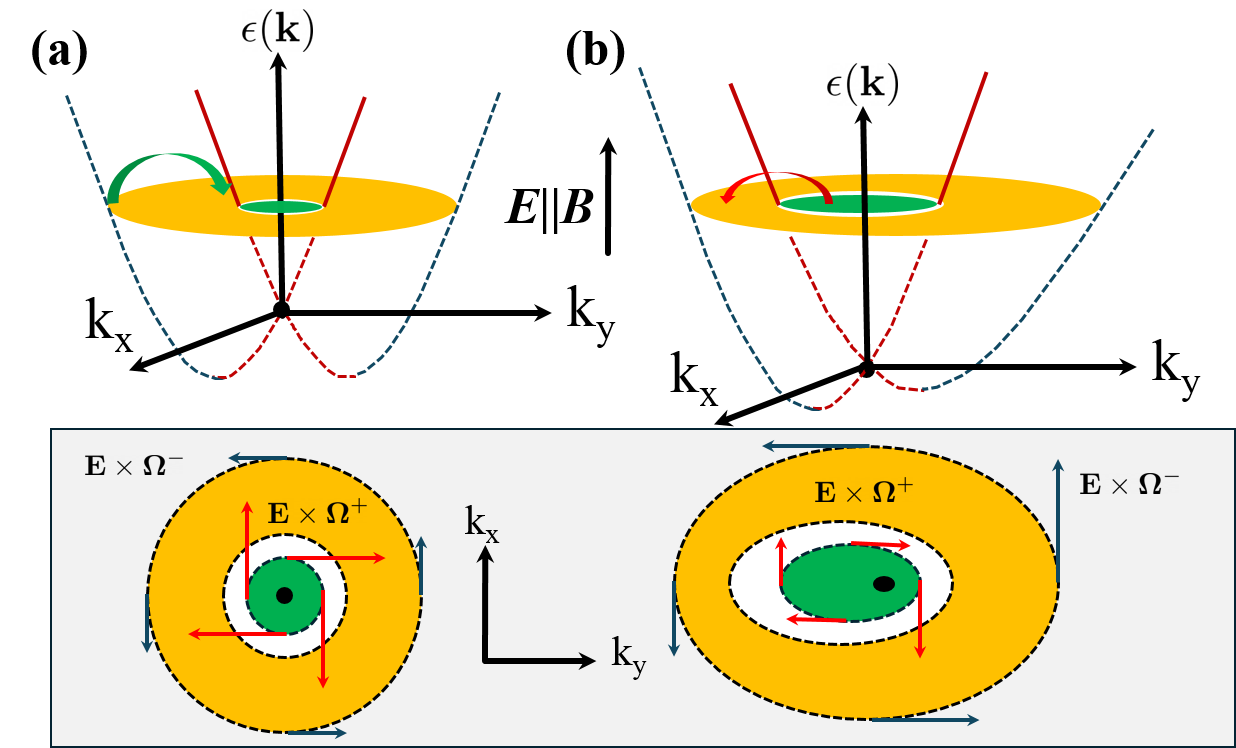}
    \caption{Schematic representation of the physical mechanism underlying the chiral anomaly-induced nonlinear Hall effect (CNLH) in spin-orbit coupled noncentrosymmetric metals (SOC-NCMs). The two electronic bands carrying opposite Berry curvature fluxes are depicted by the yellow and green regions. Interband and intraband scattering processes are indicated by red and green arrows, respectively. (a) In the case of an untilted band dispersion, the Fermi surface in the $k_x$–$k_y$ plane retains rotational symmetry. However, due to an imbalance in the Fermi surface areas associated with the two bands, the net CNLH response remains finite and is governed by the band exhibiting the dominant contribution in magnitude and sign. (b) Introduction of tilt in the band dispersion results in a distortion of the Fermi surface, which leads to an enhancement in the overall magnitude of the CNLH response.}
\label{Origine_of_CNLH}
\end{figure}

For the case of inversion symmetry broken WSM (IWSMs) having a minimum of four nodes, one can get the imbalance of chiral charge density on nodes by adding a term in the Hamiltonian, as suggested in Ref.~\cite{PhysRevB.103.045105}. A pair of Weyl nodes having the same tilt orientation is sufficient to discuss the CNLH for IWSMs, as the properties for four nodes will be qualitatively similar and will be modified by the presence of two other nodes in magnitude. For WSMs, the CNLH is zero as long as the tilt of the Weyl cones is in opposite directions, as the individual contribution of each Weyl node gets canceled for opposite tilt orientation. For the case of SOC-NCMs, we interestingly get a non-zero CNLH even without the addition of anisotropy \textit{via} tilt. This is due to the fact that the Fermi surfaces are of unequal size, and  thus individual contributions for CNLH are unequal, the summation of which leads to a non-zero net CNLH even for the untilted case, which is schematically presented in Fig.~\ref{Origine_of_CNLH}. The presence of orbital magnetic moment (OMM), tilt and spin Zeeman terms each enhanced the magnitude of CNLH. We first recover previous results: we have found that CNLH is quadratic in $B$, negative, symmetric about $B=0$, and does not show any sign reversal upon increment of interband scattering strengths. The presence of tilt in the dispersion introduced a linear in-$B$ term in the conductivity expressions, resulting in `weak and strong' sign reversals. All four scattering sites taken show the same trends with respect to the tilt along the $x$-direction. We further notice that CNLH is highly anisotropic with respect to the direction of the tilt, in the presence of all four types of impurity sites, which has never been explored before. The article is arranged as follows. Section~\ref{SOC-NCMs} is devoted to the detailed study of the SOC-NCM Hamiltonian and how directional anisotropy in the presence of tilt is incorporated in the BTE. We discuss the main results of our paper in section~\ref{r_results}. Finally we conclude our discussions in section~\ref{c_conclusion}. Remaining analytical calculations including  discussion of the quasi-classical Boltzmann transport formalism(BTE) that incorporates charge conservation and momentum-dependent scattering for magnetic as well as non-magnetic scattering sites are presented in the Appendix.\\

\section{Spin-orbit coupled (SOC) noncentrosymmetric metals (NCMs) with weak Zeeman field}
\label{SOC-NCMs}
\begin{figure}
    \centering
    \includegraphics[width=.8\columnwidth]
   {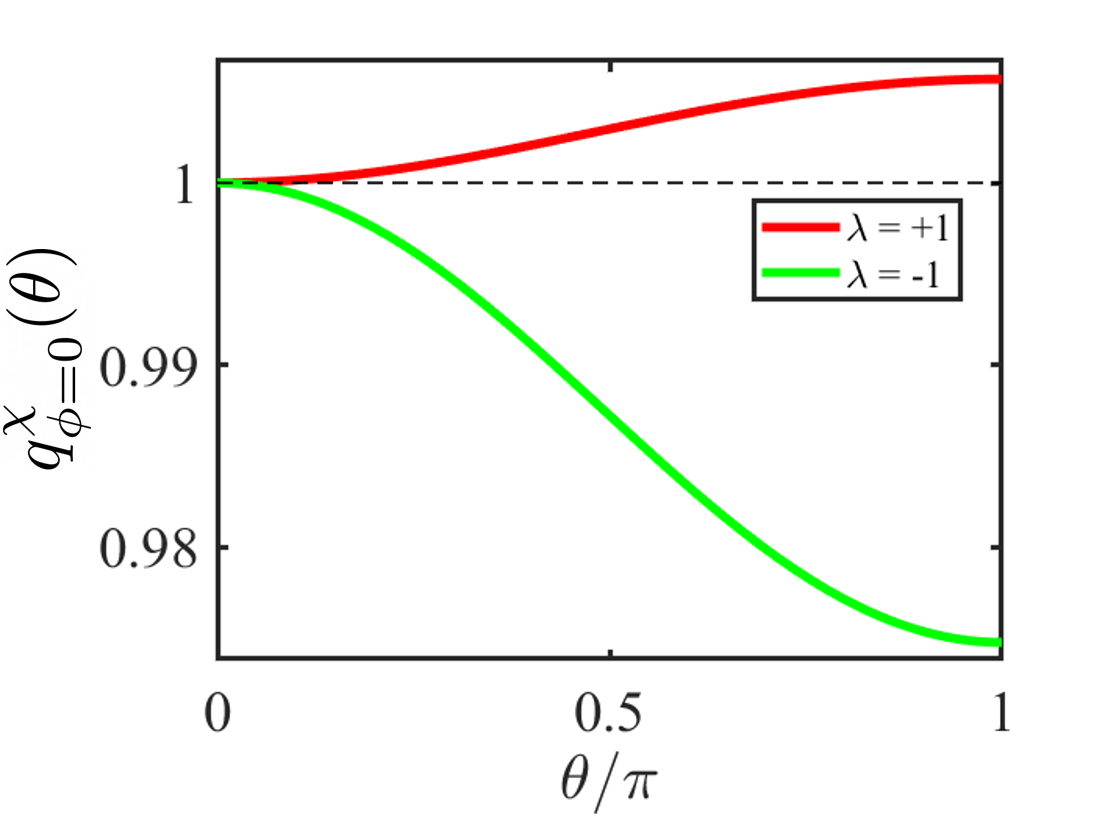}
    \caption{Plot of the normalized chiral wave vector $q^\chi$ as a function of the polar angle $\theta$ at fixed azimuthal angle $\phi=0$. The quantity $q^\chi$ is computed from Eq.~\ref{eq:E_Eigen_value_SOC_3_in_q}, incorporating the contribution of the orbital magnetic moment (OMM). The mode corresponding to chirality $\chi=+1$ exhibits a maximum at $\theta = \pi$, whereas the $\chi=-1$ mode attains a minimum at the same angular position.}
\label{fig:SOC_K1_nd_K2.png}
\end{figure}

The Hamiltonian of the SOC-NCMs having tilt and a small perturbative Zeeman Field can be written in the following form:
\begin{align}
H_{\mathrm{SOC}}(\mathbf{k}) = \frac{\hbar^2 k^2}{2 m} \sigma_{0} + \hbar \vartheta \mathbf{k} \cdot \mathbf{\sigma} + \hbar \vartheta (k_{x} t_{x} + k_{z} t_{z}) \sigma_{0} + H_{\mathrm{Z}}.
\label{eq:H_SOC_1}
\end{align}
Here, the second term accounts for the spin-orbit coupling, where $m$ denotes the effective electron mass and $\boldsymbol{\sigma}$ represents the vector of Pauli spin matrices. The parameters $t_{x,z}$ introduce anisotropy in the Hamiltonian along the $x$ and $z$ directions respectively. The Zeeman interaction is described by $H_{\mathrm{Z}} = -\boldsymbol{\mathcal{M}} \cdot \boldsymbol{\sigma}$, where the Zeeman field $\boldsymbol{\mathcal{M}}$ is related to the external magnetic field through $\boldsymbol{\mathcal{M}} = -\frac{\mu_B g}{2} \mathbf{B}$. Here, $\mu_B$ is the Bohr magneton and $g$ is the spin Landé $g$-factor (typically $g \sim 50$~\cite{yakunin2010spin,xie2021kramers}). For analytical tractability, we restrict the Zeeman field to lie within the $xz$-plane, i.e., $\boldsymbol{\mathcal{M}} = \mathcal{M} (\cos{\gamma}, 0, \sin{\gamma})$. Consequently, Eq.~\ref{eq:H_SOC_1} takes the form:
\begin{align}
    H_{\mathrm{SOC}}(\mathbf{k}) = \frac{\hbar^2 k^2}{2 m} \sigma_{0} &+ \hbar \vartheta \left( \mathbf{k - \frac{\boldsymbol{\mathcal{M}}}{\hbar \vartheta}} \right) \cdot \boldsymbol{\sigma} \nonumber\\
    &+\hbar \vartheta (k_{x} t_{x} + k_{z} t_{z}) \sigma_{0}.
\label{eq:H_SOC_2}
\end{align}
To make the effect of Zeeman field more transparent we employ the following coordinate transformation: $\mathbf{k}_{F} = \mathbf{h} + \mathbf{q}_{F}$, where, 
 $\mathbf{h} = \frac{\boldsymbol{\mathcal{M}}}{\hbar \vartheta}$. So, in the new coordinates Eq.~\ref{eq:H_SOC_2} takes the following form:
 \begin{align}
    H_{\mathrm{SOC}}(\mathbf{q}) = \frac{\hbar^2 q^2}{2 m} \sigma_{0} &+ \hbar \vartheta \mathbf{q} \cdot \boldsymbol{\sigma} \nonumber\\
    &+\hbar \vartheta (q_{x} t_{x}^{\prime} + q_{z} t_{z}^{\prime}) \sigma_{0} + H_{\mathrm{0}},
\label{eq:H_SOC_3_in_q}
\end{align}
where, \( t_{x}^{\prime} = \left( t_{x} + \frac{\hbar^2}{2m} \frac{2 \mathcal{M} \cos{\gamma}}{(\hbar \vartheta)^2} \right) \), \( t_{z}^{\prime} = \left( t_{z} + \frac{\hbar^2}{2m} \frac{2 \mathcal{M} \sin{\gamma}}{(\hbar \vartheta)^2} \right) \), and the constant energy shift is given by \( H_{0} = \left( \frac{\mathcal{M}}{\hbar \vartheta} \right)^2 \frac{\hbar^2}{2m} + t_{z} \mathcal{M} \sin{\gamma} + t_{x} \mathcal{M} \cos{\gamma} \). Thus, under a coordinate transformation that shifts the origin, the Zeeman field effectively renormalizes the tilt components along the \(x\)- and \(z\)-directions as \( t^{\mathrm{Z}}_{x} = \frac{\hbar^2}{2m} \frac{2 \mathcal{M} \cos{\gamma}}{(\hbar \vartheta)^2} \) and \( t^{\mathrm{Z}}_{z} = \frac{\hbar^2}{2m} \frac{2 \mathcal{M} \sin{\gamma}}{(\hbar \vartheta)^2} \), respectively. Additionally, it introduces a tilt-dependent constant term to the Hamiltonian. This constant shift alters the Fermi surface profile as a function of the angle \(\gamma\). Equation~\ref{eq:H_SOC_3_in_q} retains the structural form of the Hamiltonian used for SOC-NCMs in Ref.~\cite{varma2024magnetotransport}, allowing for straightforward diagonalization. The resulting energy eigenvalues are given by:
\begin{align}
    \epsilon_{\mathrm{SOC}}(q) &= \frac{\hbar^2 q^2}{2 m} + \chi \hbar \vartheta \mathbf{q} \nonumber\\
    &+\hbar \vartheta (q_{x} t_{x}^{\prime} + q_{z} t_{z}^{\prime}) + H_{\mathrm{0}},
\label{eq:E_Eigen_value_SOC_3_in_q}
\end{align}
where $\chi = \pm1$ denotes the two spin-orbit split bands. To obtain a constant energy contour—analogous to the case of Weyl semimetals—one must invert Eq.~\ref{eq:E_Eigen_value_SOC_3_in_q} with respect to $q$. In the transformed coordinate system, the expression for the orbital magnetic moment (OMM) retains a form similar to that of Weyl semimetals in the original coordinates. The inclusion of the OMM term renders Eq.~\ref{eq:E_Eigen_value_SOC_3_in_q} a cubic equation in $q$, yielding three possible solutions for each value of $\chi$. Among the three possible solutions, only one is physically acceptable, taking into consideration the validity of our low-energy model. The two other roots are not well behaved as $\mathbf{B}\to 0$. Moreover, the other ``spurious" roots have a momentum range far from the low-energy window where our model does not hold. The physical solution for $q$ selected for analysis is plotted as a function of $\theta$ at $\phi = 0$ in Fig.~\ref{fig:SOC_K1_nd_K2.png} for both $\chi = \pm 1$. The Fermi surface undergoes a chirality-dependent energy shift due to the OMM, i.e., $\epsilon^\chi_\mathbf{k} \rightarrow \epsilon_\mathbf{k} - \mathbf{m}^\chi_\mathbf{k} \cdot \mathbf{B}$.
As a consequence of the modified dispersion relation, the velocity components are also altered and are given by the following expressions:

\begin{align}
v^{\chi}_x&= \left(\frac{\hbar}{m}\right) k_{x} +\vartheta\frac{k_x}{k}+v_Ft^{\chi}_x\nonumber\\
&+\frac{u^{\chi}_{2}}{k^2} \left(\cos{\gamma}\left(1-\frac{2k^2_x}{k^2}\right)+\sin{\gamma}\left(\frac{-2k_x k_z}{k^2}\right)\right), \nonumber\\
v^{\chi}_y&= \left(\frac{\hbar}{m}\right) k_{y} + \vartheta\frac{k_y}{k}\nonumber\\ 
&+\frac{u^{\chi}_{2}}{k^2}\left(\cos{\gamma}\left(\frac{-2k_x k_y}{k^2}\right)+\sin{\gamma}\left(\frac{-2k_y k_z}{k^2}\right)\right),\nonumber\\
v^{\chi}_z&= \left(\frac{\hbar}{m}\right) k_{z} + \vartheta\frac{k_z}{k}+v_Ft^{\chi}_z\nonumber\\ 
&+\frac{u^{\chi}_{2}}{k^2}\left(\cos{\gamma}\left(\frac{-2k_x k_z}{k^2}\right)+\sin{\gamma}\left(1-\frac{2k^2_z}{k^2}\right)\right)
\label{velocity_components}
\end{align}
with $u^\chi_2=\frac{\chi e \vartheta B}{2 \hbar}$.
\section{Results} \label{r_results}
Before discussing the CNLH, here we would like to briefly review the concept of different types of sign reversal. The sign of longitudinal magnetoconductivity (LMC) in Weyl materials has been the focus of extensive study in previous research. In the case of untilted Weyl semimetals, the chiral anomaly is expected to produce a positive LMC under conditions of weak interband scattering; however, with sufficiently strong interband scattering, the sign of LMC reverses~\cite{knoll2020negative,sharma2020sign,sharma2023decoupling}. Conversely, even a slight tilting of the Weyl cone can lead to negative LMC along a specific direction of the magnetic field, even in the regime of weak interband scattering. Despite the similarity in sign reversal, the underlying mechanisms in these scenarios are fundamentally distinct, giving rise to the classifications of `strong-sign-reversal' and `weak-sign-reversal,' as outlined in Ref.~\cite{ahmad2023longitudinal,varma2024magnetotransport}. We provide a brief review here. A general expression for the magnetoconductivity tensor can be expressed as~\cite{ahmad2023longitudinal}  
\begin{align}  
\sigma_{ij}(B) = \sigma_{ij}^{(0)} + (B-B_0)^2 \sigma_{ij}^{(2)},  
\label{Eq-sij-fit}  
\end{align}  
\begin{figure}
    \centering   \includegraphics[width=\columnwidth]{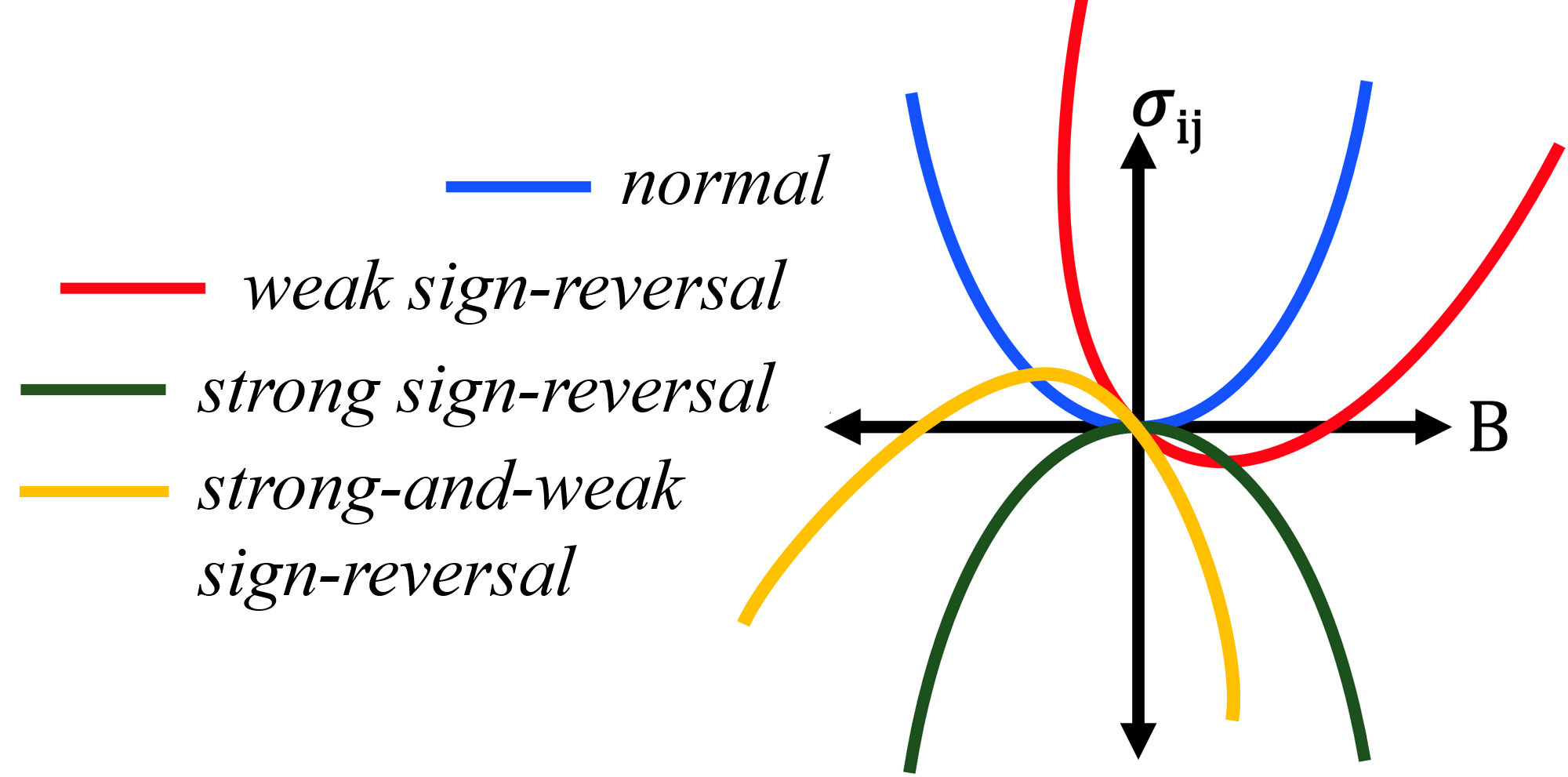}
    \caption{A schematic illustration depicting the characteristic regimes of magnetoconductivity $\sigma_{ij}$ in chiral Weyl semimetals, highlighting the phenomena of \emph{weak sign reversal}, \emph{strong sign reversal}, and the coexisting regime of \emph{strong-and-weak sign reversal}, in contrast to the conventional quadratic-in-$B$ magnetoconductivity response~\cite{ahmad2023longitudinal,varma2024magnetotransport}.}   \label{fig:signreverseschematic}
\end{figure}
which integrates (i) normal quadratic $B$-dependence, (ii) linear-in-$B$ dependence with sign change along a specific magnetic field direction, and (iii) quadratic-in-$B$ dependence with a negative sign, within a unified framework.  
The features distinguishing `weak-sign-reversal' include (i) $B_0 \neq 0$, (ii) $\sigma_{ij}^{(0)} \neq \sigma_{ij}(B=0)$, and (iii) $\mathrm{sign }\; \sigma_{ij}^{(2)}>0$. In this situation, the vertex of the magnetoconductivity parabola shifts away from the origin, and the conductivity exhibits opposite signs for small positive and negative magnetic fields. However, the parabola's orientation remains positive, i.e., $\mathrm{sign }\; (\sigma_{ij}^{(2)})>0$.  
In contrast, `strong-sign-reversal' is identified by $\mathrm{sign }\; (\sigma_{ij}^{(2)})<0$, which signifies a complete inversion of the parabola's orientation. Tilting of Weyl cones can result in `weak-sign-reversal,' while strong interband scattering or strain typically leads to `strong-sign-reversal'~\cite{ahmad2023longitudinal,varma2024magnetotransport}. Remaining possibility, i.e., completely reverse and shifted parabola results in `strong and weak-sign-reversal'. Fig.~\ref{fig:signreverseschematic} illustrates the distinction between these three cases schematically.
\begin{figure}
    \centering  \includegraphics[width=.95\columnwidth]
    {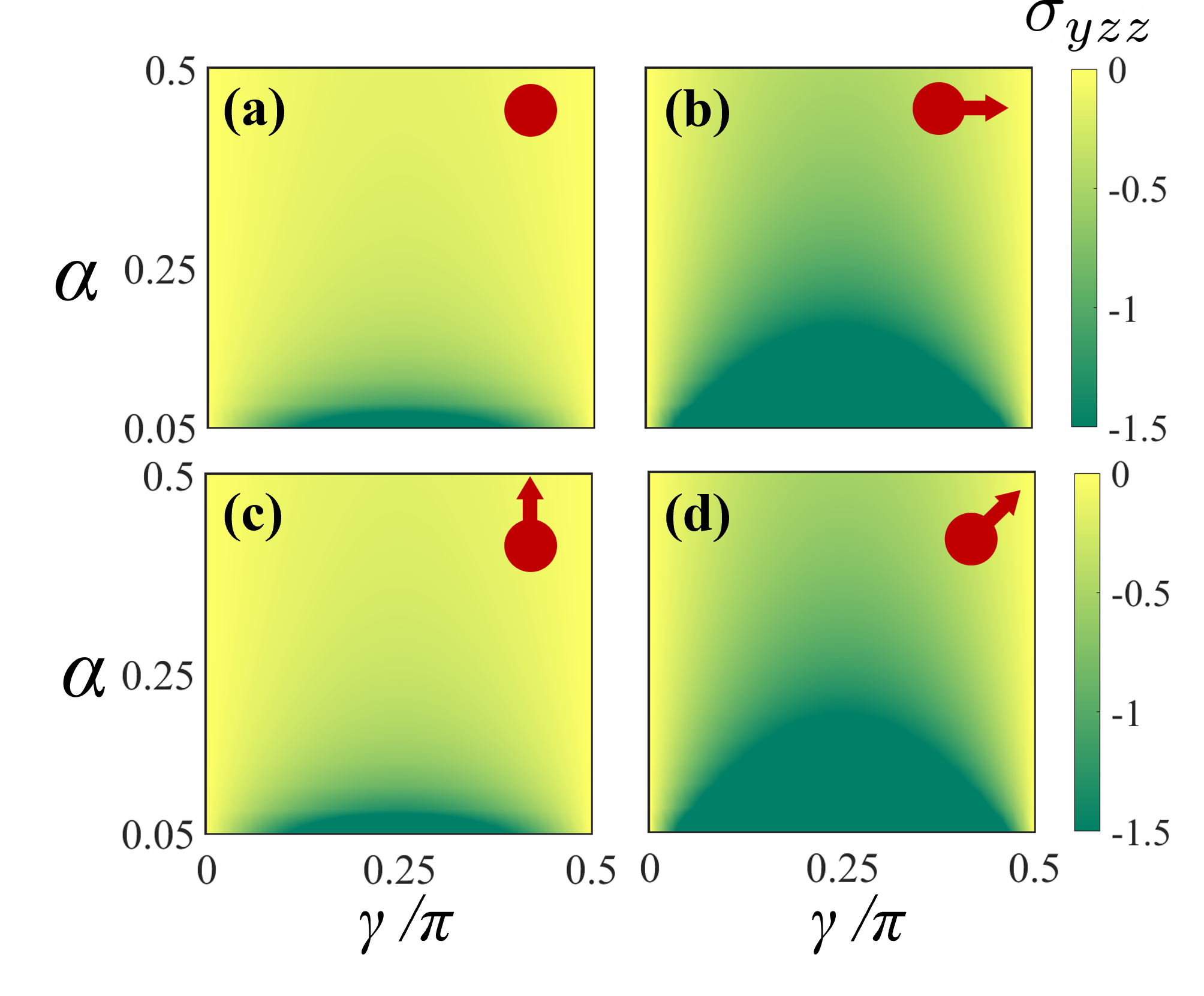}
    \caption{The chiral anomaly-induced nonlinear Hall response (CNLH) is analyzed as a function of the interband scattering strength $\alpha$ and the angle $\gamma$, at a fixed magnetic field strength of $B = 0.50~\mathrm{T}$. Panel (a) corresponds to the case of point-like non-magnetic impurity scattering, while panels (b), (c), and (d) represent magnetic impurity scattering with spin orientation aligned along the $\hat{x}$-, $\hat{y}$-, and $\hat{z}$-directions, respectively. The omission of the orbital magnetic moment (OMM) in our analysis affects only the overall amplitude—typically leading to a reduction—without altering the qualitative behavior. Hence, OMM has been included in all the present discussion.}
\label{fig:ZF1_OMM1_t0_B0p5.png}
\end{figure}
\subsection{CNLH in absence of tilt}
In the absence of tilt in the band dispersion in SOC-NCMs, the chiral anomaly-induced nonlinear Hall (CNLH) conductance exhibits a quadratic dependence on the magnetic field $B$ and remains symmetric about $B = 0$. In contrast to the linear-response regime—where a positive longitudinal magnetoconductance is typically observed due to the chiral anomaly—we consistently observe a \textit{negative} magnetoconductance in the nonlinear regime. This behavior, characterized by a sign opposite to the linear response, is referred to here as a \textit{strong sign reversal}. To examine the robustness of this feature, we analyze the behavior of CNLH conductance under varying strengths of interband scattering, denoted by $\alpha$. As expected, the magnitude of negative conductance decreases (i.e., becomes less negative) with increasing $\alpha$, indicating a suppression of anomaly-induced transport due to enhanced decoherence between bands. As established in earlier studies, the dependence of the conductance on the angle $\gamma$ between the magnetic field $\mathbf{B}$ and increasing $x$-direction shows that CNLH vanishes at $\gamma = 0$ and $\gamma = \pi/2$, reaching a maximum at $\gamma = \pi/4$\cite{ahmad2025chiral}. The $\gamma=\pi/2$ corresponds to the case when $\mathbf{E||B}$. The chiral anomaly source term is proportional to ($\mathbf{E}\cdot\mathbf{B}$), and is maximal for $\mathbf{E||B}$. However, the \textit{direction} of current is controlled by the anomalous velocity $\mathbf{v_a} \propto ( \mathbf{E} \times \Omega^\chi)$, and by the remaining angular dependence of the integrand.
For the untilted SOC-NCM (and in the absence of additional anisotropy in the scattering), the system has exact rotational symmetry in the $k_x-k_y$ plane when $\mathbf{B}$ is parallel to $\mathbf{E}$. The Berry curvature has a hedgehog-like structure, and the combination of $\mathbf{v_a}$, the distribution function, and the scattering kernel is such that the integrand for $J_y$ is odd under $\phi \to \phi+\pi$. As a result, the y-component of CNLH integrates to zero. 
Intuitively, when $\mathbf{E||B}$, the anomaly pumps charge between the Fermi surfaces along the field direction, but the induced transverse current (along the $y$-direction in our case) cancels pairwise due to the residual cylindrical symmetry. There is no preferred ``Hall axis" in the plane perpendicular to $\mathbf{E}$, and hence the net CNLH $J_y$ must vanish. At $\gamma = 0$, the projection $\mathbf{E} \cdot \mathbf{B}$ associated with the chiral anomaly becomes zero, thereby eliminating the anomaly-induced transport response. Furthermore, we investigate the influence of impurity type on the magnitude and profile of the CNLH conductance. We consider four distinct impurity scattering profiles: non-magnetic ($\sigma_0$) and magnetic impurities polarized along orthogonal directions ($\sigma_x$, $\sigma_y$, and $\sigma_z$). Although the overall qualitative behavior of the CNLH signal remains consistent across these impurity types, we observe significant quantitative variations. In particular, magnetic impurity types $\sigma_x$ and $\sigma_y$ lead to a larger magnitude of conductance, as illustrated in Fig.~\ref{fig:ZF1_OMM1_t0_B0p5.png}, whereas $\sigma_0$ and $\sigma_z$ impurities yield comparatively smaller conductance amplitudes.

The observed transport behavior can be understood through an analysis of the angular dependence of the scattering amplitudes associated with different impurity types. Specifically, from the calculated scattering probabilities (for the case $\chi\neq \chi^\prime$,~refer Fig.~\ref{Fig:Backscattering_Schematic}(b)) we find that for non-magnetic ($\sigma_0$) and longitudinal magnetic ($\sigma_z$) impurities, the interband forward scattering processes (i.e., scattering angle $\theta = 0$) are significantly suppressed, while the interband back scattering ($\theta = \pi$) is enhanced. This enhanced back scattering leads to a reduction in net conductivity, as charge carriers experience stronger momentum-relaxing collisions in the transport direction. In contrast, for transverse magnetic impurities, characterized by $\sigma_x$ and $\sigma_y$, the situation is reversed. These impurity types favor enhanced forward inter-band scattering while suppressing back scattering. Consequently, charge carriers propagating along the forward direction undergo fewer momentum relaxing collisions, resulting in an enhancement of the overall conductance.
\begin{figure}
    \centering
    \includegraphics[width = .98\columnwidth]
    {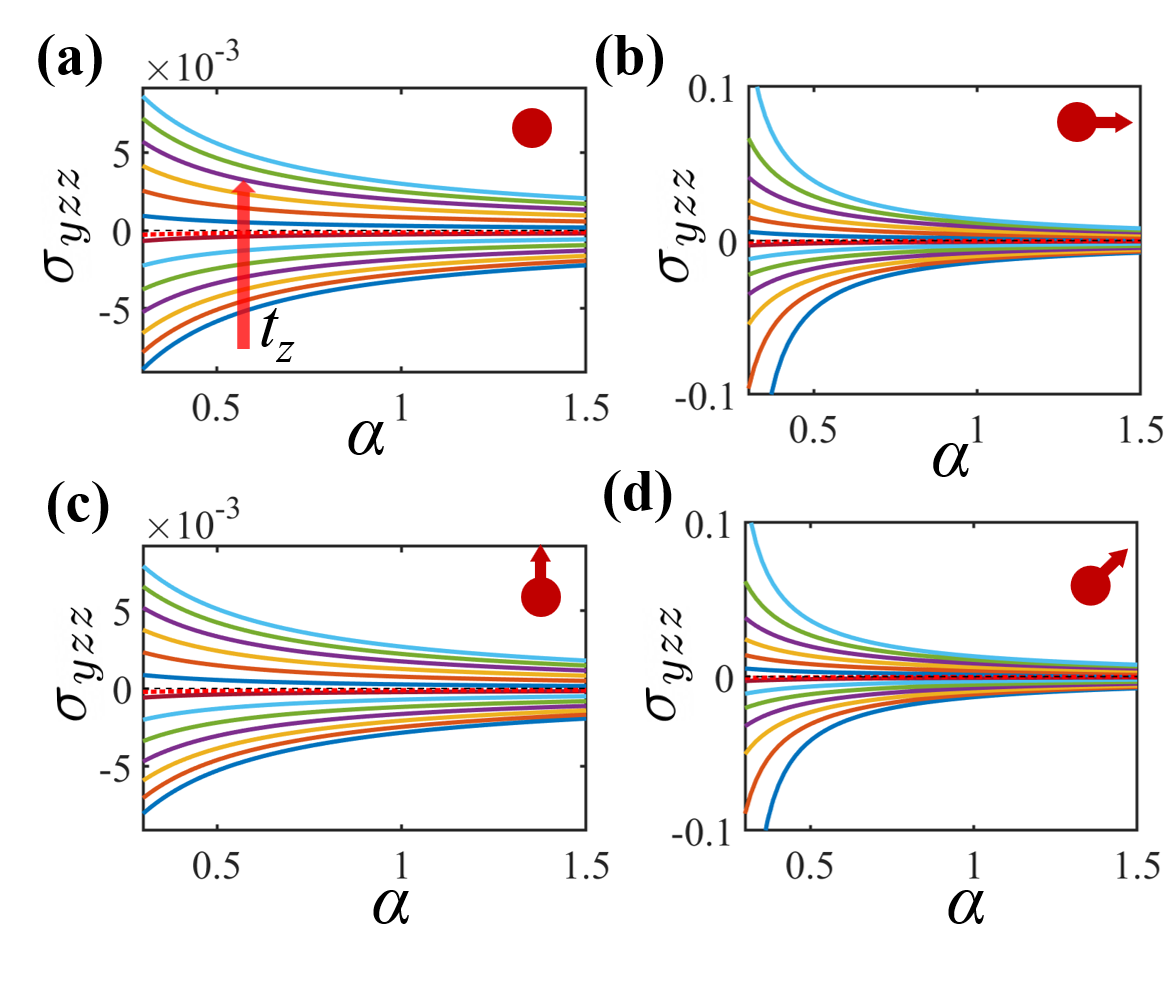}
    \caption{The chiral anomaly-induced nonlinear Hall conductivity (CNLH) is analyzed as a function of the dimensionless interband scattering strength \((\alpha)\), for various values of the tilt parameter \(t_z\), under a fixed magnetic field \(B = 0.50~\text{T}\) oriented along the direction \(\gamma = \pi/4\). The arrows indicate the progression of \(t_z\) from \(-0.98\,v_F\) to \(+0.98\,v_F\), illustrating the effect of tilt modulation on the nonlinear transport response.}
\label{fig:CNLH_alp_tz_vary.png}
\end{figure}

\begin{figure}
    \centering
    \includegraphics[width=.95\columnwidth]
    {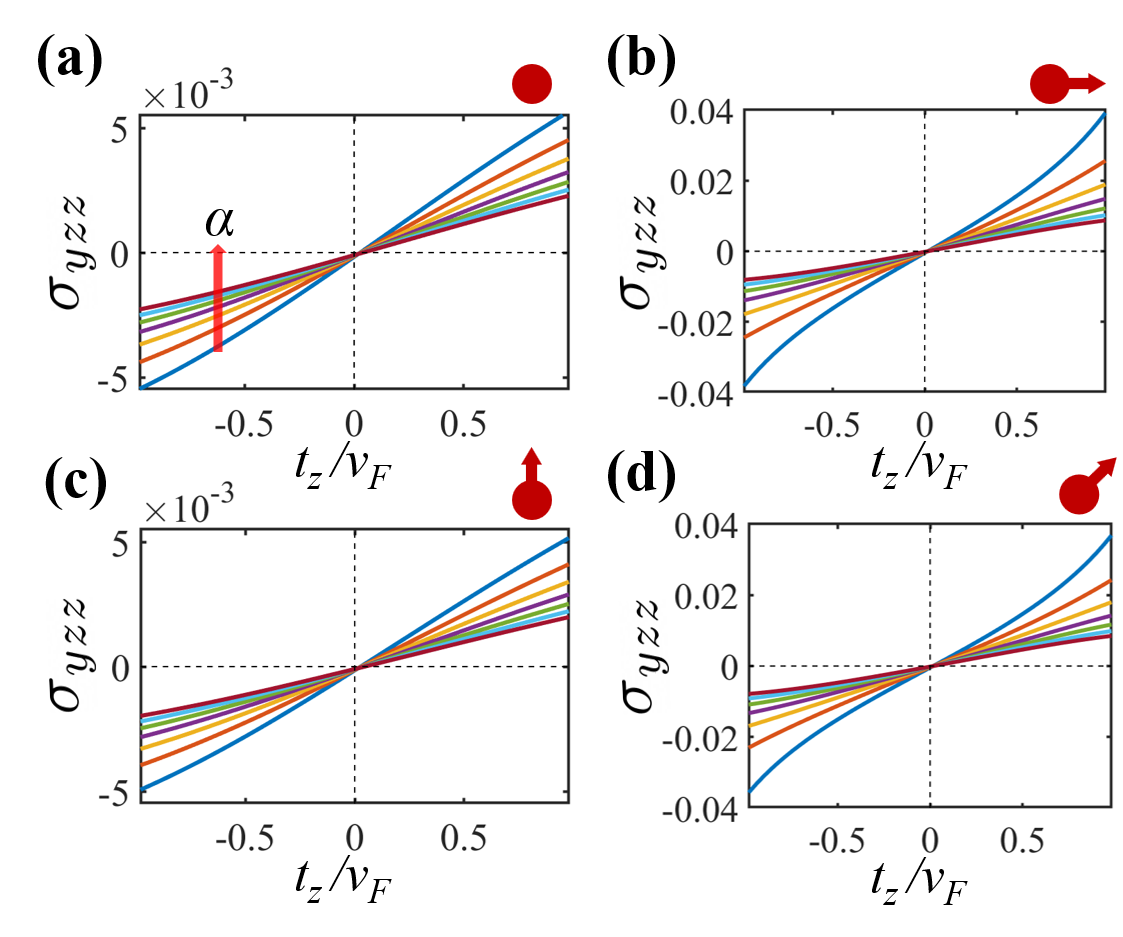}
    \caption{The chiral anomaly-induced nonlinear Hall response (CNLH) is analyzed as a function of tilt along the $z$-axis ($t_z$) for various values of interband scattering strength ($\alpha$), under a fixed magnetic field $B = 0.50~\text{T}$ oriented at an angle $\gamma = \pi/4$. Similar to the case with finite $t_x$, the CNLH vanishes at a finite, nonzero value of the tilt parameter $t_z$, indicating a symmetry-breaking threshold. For all four impurity scattering configurations considered—encompassing scalar and magnetic disorder profiles—the CNLH exhibits an approximately linear and monotonic dependence on $t_z$ within the parameter regime explored.}
\label{fig:CNLH_tz_alp_vary.png}
\end{figure}
\subsection{CNLH in the presence of tilt along the z-direction}
In this analysis, we introduce a finite tilt component along the $z$-axis, denoted by $t_z$, and investigate the dependence of the chiral anomaly-induced nonlinear Hall (CNLH) conductance on both the interband scattering strength $\alpha$ and the tilt parameter. Figure~\ref{fig:CNLH_alp_tz_vary.png} illustrates the variation of the CNLH conductance as a function of $\alpha$ for several values of $t_z$. Our results indicate that the CNLH response can assume both positive and negative values, contingent on the sign and magnitude of the tilt $t_z$, highlighting its directional sensitivity. As anticipated, the overall magnitude of the conductance exhibits a monotonic decrease with increasing $\alpha$, showing a rapid decay in the low-$\alpha$ regime and approaching a saturation limit asymptotically for large $\alpha$ values. To ensure comprehensive coverage, we have performed the analysis for all four types of impurity scatterers: non-magnetic ($\sigma_0$) and magnetic ($\sigma_x$, $\sigma_y$, $\sigma_z$). Consistent with the zero-tilt scenario, we observe that the conductance magnitudes associated with $\sigma_x$ and $\sigma_y$ magnetic scatterers remain significantly larger than those corresponding to $\sigma_0$ and $\sigma_z$. This enhancement is attributed to the suppression of interband back scattering by the spin-flipping scatterers ($\sigma_x,\sigma_y$), as discussed previously. Furthermore, the pronounced suppression of conductance with increasing $\alpha$ in the $\sigma_x$ and $\sigma_y$ cases may be attributed to the increased phase-space volume for scattering introduced by the tilt-induced anisotropy along the $z$-direction, which enhances the density of accessible final states in momentum space.

We further analyze the dependence of the CNLH on the tilt parameter $t_z$ for various values of the interband scattering strength $\alpha$, as illustrated in Fig.~\ref{fig:CNLH_tz_alp_vary.png}. The conductance profile displays a linear behavior with respect to $t_z$, characterized by a predominantly linear trend for all 4 cases . As expected, impurity-induced scattering through transverse magnetic scatterers $(\sigma_x, \sigma_y)$ yields a significantly enhanced conductance compared to that arising in the presence of non-magnetic and longitudinal magnetic scatterers $(\sigma_0, \sigma_z)$. Interestingly, for the $(\sigma_x, \sigma_y)$ scatterers, a mild deviation from strict linearity is discernible, while the symmetry about $t_z = 0$ remains preserved. This feature is indicative of the nature of tilt-induced symmetry breaking in the low-energy SOC-NCM Hamiltonian. \\

\subsection{CNLH in the presence of tilt along the x-direction}
\begin{figure*}
    \centering
    \includegraphics[width = 1.98\columnwidth]
    {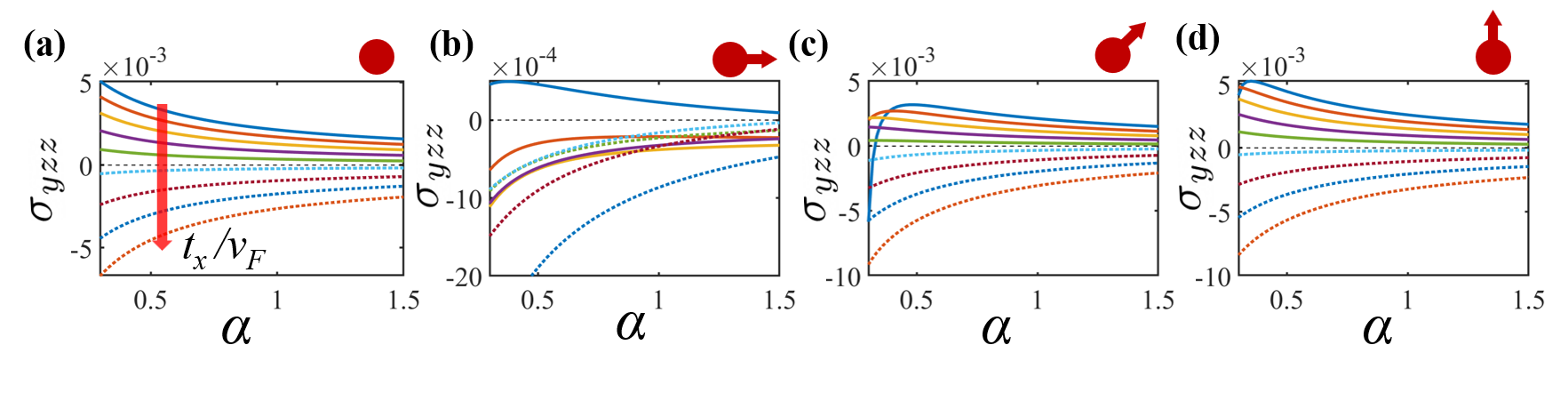}
    \caption{The chiral anomaly-induced nonlinear Hall conductivity (CNLH) is plotted as a function of the dimensionless interband scattering strength parameter ($\alpha$) for various values of the tilt parameter ($t_x$), under a fixed magnetic field $B = 0.50~\text{T}$ applied at an oblique angle characterized by $\gamma = \pi/4$ with respect to $\hat{x}$. The dashed curves correspond to negative tilt configurations, i.e., $t_x/v_F < 0$. Panels (a) and (d) exhibit qualitatively similar CNLH responses, whereas panels (b) and (c) reveal distinct deviations due to the interplay between the tilt-induced anisotropy and interband scattering.}
\label{fig:CNLH_alp_tx_vary.png}
\end{figure*}

\begin{figure*}
    \centering
    \includegraphics[width = 1.95\columnwidth]
    {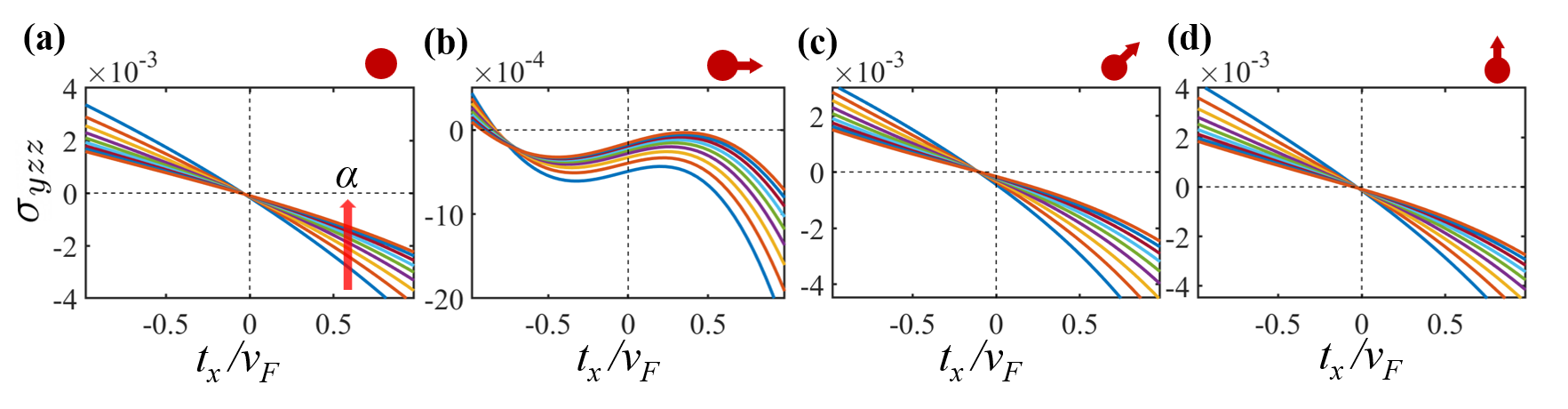}
    \caption{The chiral anomaly-induced nonlinear Hall (CNLH) conductivity is plotted as a function of the tilt parameter ($t_x$) for various interband scattering strengths ($\alpha$), under a fixed magnetic field magnitude $B = 0.50~\mathrm{T}$ oriented along the direction $\gamma = \pi/4$. Notably, the CNLH response vanishes at a finite (non-zero) value of tilt, indicating a nontrivial cancellation mechanism. With the exception of panel (b), the CNLH exhibits a predominantly linear and monotonic dependence on $t_x$, reflecting the symmetry-breaking role of tilt in modulating the anomalous transport.}
\label{fig:CNLH_tx_alp_vary.png}
\end{figure*} 
\begin{figure*}
    \centering
    \includegraphics[width = 1.95\columnwidth]
    {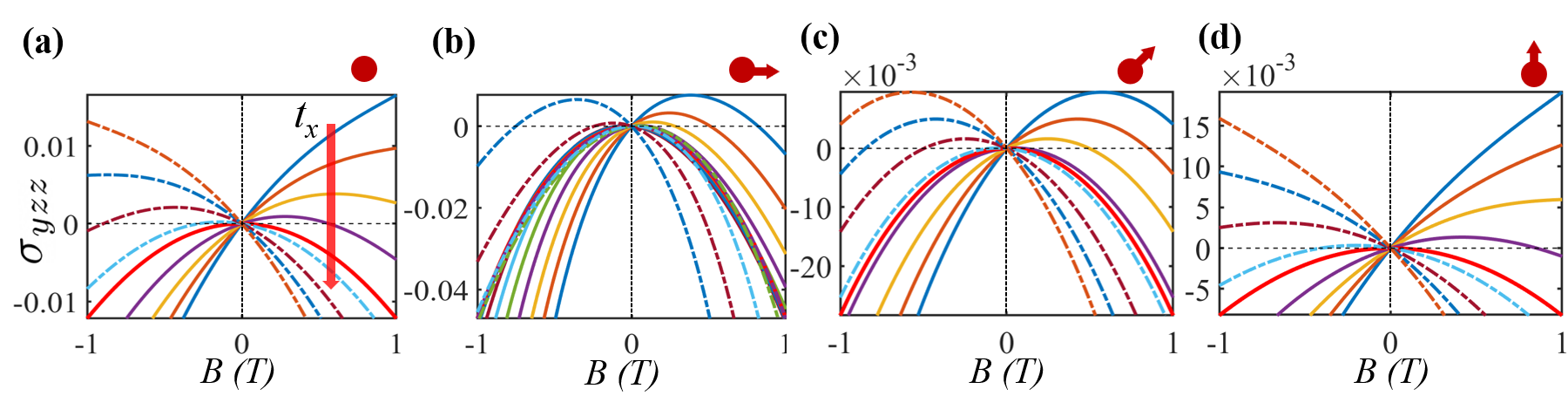}
    \caption{The chiral anomaly-induced nonlinear Hall (CNLH) response is plotted as a function of magnetic field $B$ for varying tilt strengths ($t_x$) at a fixed interband scattering parameter ($\alpha = 0.30$). In the presence of finite tilt, the CNLH curve exhibits an inverted parabolic profile that is displaced from the origin ($B = 0$), signaling the emergence of a \emph{strong-and-weak sign reversal} regime. Note that the legends are not uniform across all subplots. The non-monotonic dependence of the CNLH amplitude on $t_x$ in the presence of a magnetic impurity aligned along the $x$-axis is distinctly illustrated in panel (b).}
\label{fig:CNLH_vs_B_tx_vary_alp_0p30.png}
\end{figure*}
Here, we discuss our results when a non zero tilt is considered in the x-direction. As studied previously, we have plotted conductance as a function of $\alpha$ for different values of the tilt parameter $t_x$, as shown in Fig.~\ref{fig:CNLH_alp_tx_vary.png}. Here, we see an interesting result. We have found that there is an asymmetry in the conductance profile with respect to $t_x$ for the case of $(\sigma_x,\sigma_y)$ impurity scatterers. Moreover, the conductance is lesser in magnitude, particularly for the $\sigma_x$ impurity scatterer. A similar trend is followed in Fig.~\ref{fig:CNLH_tx_alp_vary.png}. We found that the conductance profile looks asymmetric and non-monotonous in addition to being lesser in magnitude, for the $\sigma_x$ impurity scatterer. The reason for this can be traced down to a combination of several factors. The tilt along the x-direction increases the available number of states in the Fermi surface, along that direction. Moreover, $\sigma_x$ magnetic impurity has a large probability of scattering fermions along the positive x-direction. The larger availability of momentum states along with the larger probability of scattering in this preferred direction aided by $\sigma_x$ magnetic scatterer, makes the fermions scatter more leading to decrease in net conductance. Moreover, the conductance profile as a function of $t_x$ looks opposite compared to the $t_z$ case. This can be explained as follows. The dependence of CNLH conductance on tilt is found to be of the form $(\mathbf{t}\times \mathbf{B})\cdot (\mathbf{E}\times\mathbf{\Omega})$. For a given direction of $\mathbf{B}$, $t_x$ and $t_z$ gives opposite signs for the conductance, based on the above expression.

We have also plotted CNLH conductance as a function of the magnetic field as shown in Fig.~\ref{fig:CNLH_vs_B_tx_vary_alp_0p30.png}. Here the bright red curve represents CNLH conductance for zero tilt which shows a \textit{strong sign reversal}. However, for the case of finite tilt $t_x$, we have found that the parabola gets displaced from the origin ($B=0$) showing a positive conductance for a certain range of magnetic field. This belongs to the case represented by the yellow curve in Fig.~\ref{fig:signreverseschematic}, which we refer to as \textit{weak and strong sign reversal}. This as studied in previous works, is due to the fact that tilt adds a linear in $B$ term to the conductance. Additionally, we have also found that for the case of non-magnetic and longitudinal magnetic scatterers ($\sigma_0$,$\sigma_z$), the CNLH magnitudes are less and show larger $t_x$ dependence. Transverse magnetic scatterers ($\sigma_x$,$\sigma_y$) cause enhanced conductance with relatively less $t_x$ dependence. Interestingly, we also note that for the case of $\sigma_x$ magnetic scatterers, there is a large anisotropy in the CNLH conductance profile, which is not present for the $t_z$ case. This again is possibly due to the complex interplay between favourable scattering probabilities and extra availability of scattering states aided by $t_x$.
\section{Conclusion} \label{c_conclusion}
In this work, we have presented a detailed semiclassical analysis of the chiral anomaly-induced nonlinear Hall (CNLH) conductance in SOC-NCMs in the presence of different impurity scatterers and tilt configurations. In the absence of tilt, the CNLH conductance displays a negative quadratic dependence on the magnetic field and exhibits a strong sign reversal irrespective of the interband scattering strength unlike its linear counterpart which shows a consistent positive magnetoconductance. This behavior is found to be sensitive to the interband scattering strength $\alpha$, which suppresses the anomaly-induced transport as $\alpha$ increases. We examined the angular dependence of the CNLH conductance and its variation across four distinct impurity types—non-magnetic ($\sigma_0$) and magnetic ($\sigma_x$, $\sigma_y$, $\sigma_z$). While the qualitative trends remain robust, magnetic impurities with transverse polarization ($\sigma_x$, $\sigma_y$) produce significantly larger conductance magnitudes due to their suppressed interband back scattering. In contrast, non-magnetic and longitudinal magnetic impurities lead to enhanced interband back scattering that causes a decrease in the net conductance.

\begin{figure}
    \centering
    \includegraphics[width=\columnwidth]
    {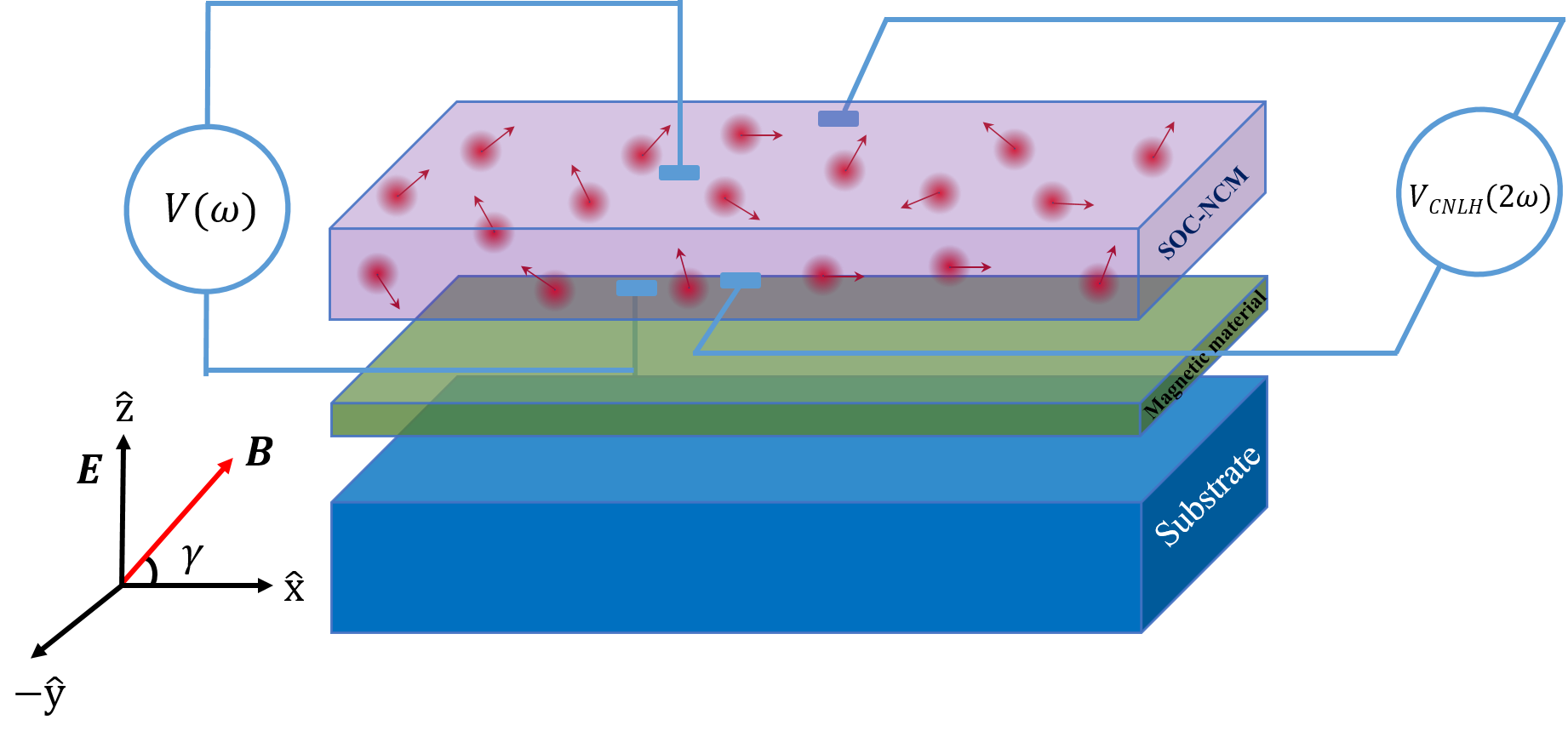}
    \caption{Experimental setup to test the  CNLH in SOC-NCMs having magnetic as well as non-magnetic scattering sites. The bottom layer represents the substrate while the middle layer is that of a magnetic material that controls the spin orientation of the magnetic scattering sites, represented by red in the top layers. }
\label{fig:Experimental_setup}
\end{figure}

Introducing a finite tilt component along either the $z$- or $x$-direction reveals rich transport behavior. In the $z$-tilted case, the CNLH conductance becomes highly anisotropic and sensitive to the sign and magnitude of the tilt parameter $t_z$. The conductance decays with increasing $\alpha$ for all impurity types, with transverse magnetic scatterers again showing a dominant response. The presence of tilt enlarges the phase-space volume for scattering, which contributes to the enhanced suppression observed in tilted scenarios. Additionally, tilt along the $x$-direction leads to an asymmetric and non-monotonic conductance profile, particularly pronounced for the $\sigma_x$ impurity type. This asymmetry arises due to the interplay between the anisotropic deformation of the Fermi surface and the preferential scattering direction associated with the $\sigma_x$ magnetic impurity. Together, these findings provide a comprehensive understanding of how tilt and impurity scattering mechanisms govern the CNLH response in Weyl systems, highlighting the tunable and anisotropic nature of anomaly-induced nonlinear transport in topological materials. \\
 \textit{Experimental proposal:}~There are several SOC-NCM candidates hosting Kramers-Weyl Fermions, that have sizable spin-orbit coupling and broken inversion symmetry that can potentially show CNLH as discussed in this paper. An exhaustive list of materials can be found in Refs.~\cite{wieder2022topological,chang2018topological}. 
From an experimental standpoint, our results suggest several clear signatures for
detecting the chiral–anomaly–induced nonlinear Hall (CNLH) response in SOC–NCMs.
An experimentally accessible indicator is the strong dependence of the CNLH
magnitude on impurity type. Our analysis shows that transverse magnetic scatterers
($\sigma_x$, $\sigma_y$) substantially enhance the CNLH response, whereas nonmagnetic or
longitudinal magnetic impurities ($\sigma_0$, $\sigma_z$) suppress it. This suggests that controlled
magnetic alloying or implantation can serve as an effective handle to amplify the nonlinear
signal and to distinguish the relative roles of intraband and interband scattering channels.
We suggest an experimental setup as shown in Fig.~\ref{fig:Experimental_setup}. In order to experimentally probe CNLH effect, we propose the following: (i) apply an AC electric field along the z-axis and a static magnetic field in the xz-plane (as in our theory), and detect the second-harmonic Hall voltage (the CNLH signal) along the y-direction, (ii) rotate $\mathbf{B}$ and map out the angular dependence, looking for the predicted vanishing at $\gamma= 0, \pi/2$ and the strong anisotropy with tilt, (iii) use controlled magnetic impurity doping or proximity to a magnetic material as shown in Fig.~\ref{fig:Experimental_setup} to tune the relative strength of $\sigma_0, \sigma_x, \sigma_y, \sigma_z$-type scattering channels, and observe the corresponding changes in the magnitude of CNLH, (iv) vary the carrier density (via chemical substitution) to shift the Fermi level and test the sensitivity of the effect to Fermi-surface asymmetry.The second harmonic Hall voltage refers to the Hall voltage measured at twice the driving frequency when the system is subjected to an ac electric field of the form $\bm{E}(t)=\bm{E}_0\cos(\omega t)$. Since the CNLH current is quadratic in the electric field, $J_y\propto E^2$, it naturally contains a component oscillating at frequency $2\omega$, which can be
experimentally isolated using standard lock-in techniques. One can possibly ask as to how our proposed experimental set up can be used to probe a frequency dependent Hall voltage while our calculations assume a steady state with a definite CNLH conductivity. However, we wish to clarify that the assumption remains valid for the proposed ac measurement provided the driving frequency is low compared to the relevant scattering rates, i.e., $\omega \tau \ll 1$, where $\tau$ denotes the characteristic momentum and interband/intraband relaxation times. In this low-frequency regime, the system follows the external drive quasi-statically and relaxes continuously via impurity scattering, allowing a time-periodic steady state to be established.
\section{Acknowledgments}
The authors sincerely acknowledge Gargee Sharma for providing scientific input and offering valuable physical insights. G.V.K. acknowledges financial support from IIT Mandi HTRA. A.A. acknowledges financial support from Project IITM/ANRF-SERB/GS/495. The authors are grateful to Shubhanshu Karoliya and Arpan Gupta for insightful discussions and valuable scientific input.
\appendix
\section{Maxwell Boltzmann Transport formalism} 
\label{App_MBT}
\begin{figure}
    \centering
    \includegraphics[width=\columnwidth]
    {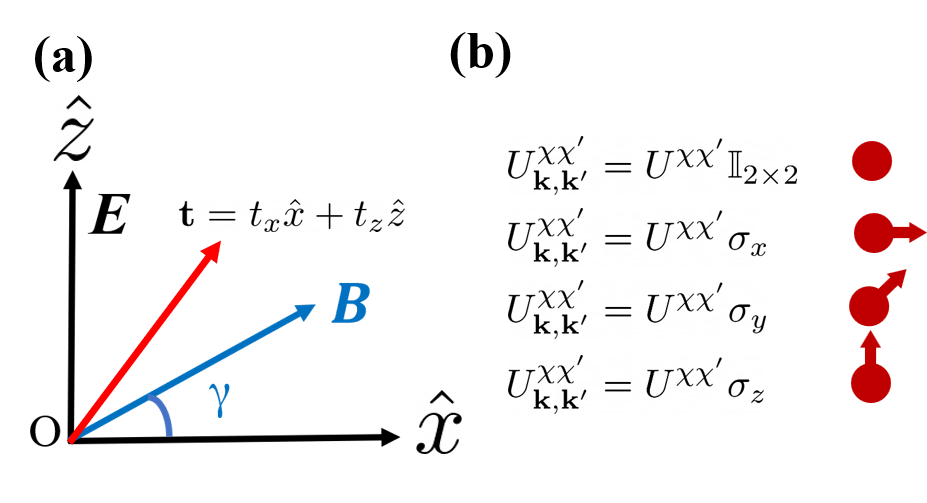}
    \caption{(a) Orientation of magnetic field $\mathbf{B}$ and tilt direction $\mathbf{t}$ in $xz$-plan. This geometry is helpful to understand the CNLH for the case of tilted dispersion along any general direction in the $xz$-plane. (b) types of impurity profile (scattering site) taken in Eq.~\ref{Fermi_gilden_rule} and symbolic representation of the same.}
\label{fig:Orientation_of_MF_and_impurity_type}
\end{figure}

The occupation of the states can be related to the distribution function $f_{\mathbf{k}}$ in phase space. Any change in equilibrium is reflected by it. We use the semiclassical Maxwell-Boltzmann equation to describe the dynamics of Weyl fermions in the presence of electric and magnetic fields. In phase space Non-equilibrium distribution function $f^{\chi}_{\mathbf{k}}$ evolves as:

\begin{align}
\dfrac{\partial f^{\chi}_{\mathbf{k}}}{\partial t}+ {\Dot{\mathbf{r^{\chi}_{\mathbf{k}}}}}\cdot \mathbf{\nabla_r}{f^{\chi}_{\mathbf{k}}}+\Dot{\mathbf{k^{\chi}}}\cdot \mathbf{\nabla_k}{f^{\chi}_{\mathbf{k}}}=I_{coll}[f^{\chi}_{\mathbf{k}}],
\label{MB_equation}
\end{align}

where $f^{\chi}_\mathbf{k} = f_{0} + g^{1, \chi}_{\mathbf{k}} + g^{2, \chi}_{\mathbf{k}} + O(E^3)$. $f_{0} $ is the standard Fermi-Dirac distribution, $g^{1,\chi}_{\mathbf{k}}$ and $g^{2,\chi}_{\mathbf{k}}$ are deviation up to the first order and second order in electric field $E$, respectively, which can be written in following form ~\cite{PhysRevB.103.045105,PhysRevB.104.205124,PhysRevB.105.125131}:
\begin{align}
g^{1,\chi}_{\mathbf{k}} = -e\left({\dfrac{\partial f_{0}}{\partial {\epsilon}}}\right)_{\epsilon=\mu}\mathbf{E}\cdot \mathbf{\Lambda^{1,\chi}_k}
\label{Eq:g1}
\end{align}
\begin{align}
g^{2,\chi}_{\mathbf{k}} = -e\left({\dfrac{\partial g^{1,\chi}_{\mathbf{k}}}{\partial {\epsilon}}}\right)_{\epsilon=\mu}\mathbf{E}\cdot \mathbf{\Lambda^{2,\chi}_k}.
\label{Eq:g2}
\end{align}
$\mathbf{\Lambda}^{1,\chi}_{\mathbf{k}}$ and $\mathbf{\Lambda}^{2,\chi}_{\mathbf{k}}$ are unknown functions (ansatz) to be evaluated. RHS in Eq.~\ref{MB_equation}, i.e., collision integral term is selected to incorporate both interband and intraband scattering, and is expressed as,
\begin{align}
 I_{coll}[f^{\chi}_{\mathbf{k}}]=\sum_{\chi' \mathbf{k}'}{\mathbf{W}^{\chi \chi'}_{\mathbf{k k'}}}{(f^{\chi'}_{\mathbf{k'}}-f^{\chi}_{\mathbf{k}})},
 \label{Collision_integral}
\end{align}
where, scattering rate ${\mathbf{W}^{\chi \chi'}_{\mathbf{k k'}}}$ calculated using Fermi's golden rule,
\begin{multline}
\mathbf{W}^{\chi \chi'}_{\mathbf{k k'}} = \frac{2\pi n}{\mathcal{V}}|\bra{u^{\chi'}(\mathbf{k'})}U^{\chi \chi'}_{\mathbf{k k'}}\ket{u^{\chi}(\mathbf{k})}|^2\\\times\delta(\epsilon^{\chi'}(\mathbf{k'})-\epsilon_F).
\label{Fermi_gilden_rule}
\end{multline}
Here, ‘\(n\)’ is the impurity concentration, ‘\(\mathcal{V}\)’ is the system volume, \(u^\chi(\mathbf{k})\) is the spinor wave function, \(U^{\chi\chi'}_{\mathbf{k}\mathbf{k}'}\) is the scattering potential profile, and \(\epsilon_F\) is the Fermi energy. We choose \(U^{\chi\chi'}_{\mathbf{k}\mathbf{k}'}\) such that it can include both magnetic and non-magnetic point-like scattering centers. In general, \(U^{\chi\chi'}_{\mathbf{k}\mathbf{k}'} = U^{\chi\chi'} \sigma_i\) with \(i = 0, x, y, z\), where \(U^{\chi\chi'}\) distinguishes the interband (\(\chi \neq \chi'\)) and intraband (\(\chi = \chi'\)) scattering. Here we would like to highlight that in our formalism, the relative strength of these two scattering channels can be tuned using the parameter defined as:
\begin{align}
\alpha =\frac{U^{\chi\chi'\neq\chi}}{U^{\chi\chi'=\chi}}.
\label{Eq: alpha}
\end{align} 
In this work, we consider the geometry represented in Fig. \ref{fig:Orientation_of_MF_and_impurity_type}(a), where the electric field is fixed along the \(z\)-direction, and the magnetic field is rotated in the \(xz\)-plane, forming an angle \(\gamma\) with the \(x\)-axis. 
Including the topological nature of the Hamiltonian, in the presence of electric ($\mathbf{E}$) and magnetic ($\mathbf{B}$) fields, semiclassical dynamics of the Bloch electrons are modified and governed by the following equation~\cite{son2012berry,knoll2020negative}:
\begin{align}
\dot{\mathbf{r}}^\chi &= \mathcal{D}^\chi \left( \frac{e}{\hbar}(\mathbf{E}\times \boldsymbol{\Omega}^\chi) + \frac{e}{\hbar}(\mathbf{v}^\chi\cdot \boldsymbol{\Omega}^\chi) \mathbf{B} + \mathbf{v}_\mathbf{k}^\chi\right) \nonumber\\
\dot{\mathbf{p}}^\chi &= -e \mathcal{D}^\chi \left( \mathbf{E} + \mathbf{v}_\mathbf{k}^\chi \times \mathbf{B} + \frac{e}{\hbar} (\mathbf{E}\cdot\mathbf{B}) \boldsymbol{\Omega}^\chi \right),
\label{Couplled_equation}
\end{align}
where, $\mathbf{p}=\hbar\mathbf{k}$, $\mathbf{v}_\mathbf{k}^\chi = \frac{1}{\hbar}\frac{\partial\epsilon^{\chi}(\mathbf{k})}{\partial\mathbf{k}}$ is band velocity, $\boldsymbol{\Omega}^\chi_\mathbf{k} = i \nabla_{\mathbf{k}} \times \langle u^{\chi}(\mathbf{k}) | \nabla_{\mathbf{k}} | u^{\chi}(\mathbf{k}) \rangle $ is the Berry curvature, and $\mathcal{D}^\chi = (1+e\mathbf{B}\cdot\boldsymbol{\Omega}^\chi/\hbar)^{-1}$ is factor by which density of states is modified due to presence of the Berry curvature. Self-rotation of the Bloch wave packet also gives rise to an orbital magnetic moment (OMM) $\mathbf{m}^{\chi}_\mathbf{k}= -\frac{ie}{2\hbar} \text{Im} \langle \nabla_{\mathbf{k}} u^{\chi}|[ \epsilon_0(\mathbf{k}) - \hat{H}^{\chi}(\mathbf{k}) ]| \nabla_{\mathbf{k}} u^{\chi}\rangle$~\cite{xiao2010berry}. To calculate $g^{i,\chi}_{\mathbf{k}}$, with $i=1~\& ~2$, we use Eq.~\ref{Couplled_equation},~\ref{Collision_integral}. Keeping terms up-to linear order in electric ($\mathbf{E}$) field, the Boltzmann transport equation can be written as,
\begin{align}
&\left[\left(\frac{\partial f_0^\chi}{\partial \epsilon^\chi_\mathbf{k}}\right) \mathbf{E}\cdot \left(\mathbf{v}^\chi_\mathbf{k} + \frac{e\mathbf{B}}{\hbar} (\boldsymbol{\Omega}^\chi\cdot \mathbf{v}^\chi_\mathbf{k}) \right)\right]\nonumber\\
 &= -\frac{1}{e \mathcal{D}^\chi}\sum\limits_{\chi'}\sum\limits_{\mathbf{k}'} W^{\chi\chi'}_{\mathbf{k}\mathbf{k}'} (g^{1,\chi}_{\mathbf{k}'} - g^{1,\chi}_\mathbf{k}).
 \label{Eq_boltz21}
\end{align}\\
Moreover, a similarly equation is also possible to write for $g^{2,\chi}_\mathbf{k}$ (i.e., second order in $\mathbf{E}$),
\begin{align}
&\left[\left(\frac{\partial g^{1,\chi}}{\partial \mathbf{k}}\right) \cdot \left(\mathbf{E} + \frac{e\mathbf{B} \cdot \mathbf{E}}{\hbar} ~\boldsymbol{\Omega}^\chi_\mathbf{k} \right)\right]\nonumber\\
 &= -\frac{\hbar}{e \mathcal{D}^\chi}\sum\limits_{\chi'}\sum\limits_{\mathbf{k}'} W^{\chi\chi'}_{\mathbf{k}\mathbf{k}'} (g^{2,\chi'}_{\mathbf{k}'} - g^{2,\chi}_\mathbf{k}).
 \label{Eq_boltz22}
\end{align} 
The present study primarily aims to examine the influence of the chiral anomaly term, while the Lorentz force contribution is neglected. This is because we work in the weak-field, weak-Lorentz-force regime where the small parameter is the cyclotron factor given by
\begin{align}
 \omega_c\tau \equiv \frac{eB}{m}\tau \ll1   
\end{align}
where $\tau$ is the relevant transport time. In this regime, the Lorentz force term in $\mathbf{\dot p}$ is suppressed by $\omega_c\tau$ compared to the electric field term. However, the chiral anomaly term $(\mathbf{E\cdot B)\Omega_k^{\chi}}$ is not suppressed in the same way. For typical SOC-NCM parameters, one has $\omega_c\tau\ll1$ in the metallic regime where the semi-classical description is valid. A large body of work is found in literature in the above mentioned metallic regime~\cite{ahmad2021longitudinal,sharma2016nernst,PhysRevB.107.115161,nandy2021chiral,li2021nonlinear,zeng2022chiral}. We fix electric field along increasing $z$-direction and magnetic field can be rotated in $xz$-plane. Therefore, $\mathbf{E} = E(0,0,1)$ and  $\mathbf{B} = B (\cos{\gamma},0,\sin{\gamma})$, i.e., for $\gamma=\pi/2$ both fields are parallel. In this case, only $z$-component of $\mathbf{\Lambda}^{\chi}_{\mathbf{k}}$ will be relevant. Eq.~\ref{Eq_boltz21} and ~\ref{Eq_boltz22} can be written to:
\begin{align}
&\mathcal{D}^{\chi}(k)\left[{v^{\chi,z}_{\mathbf{k}}}+\frac{eB\sin{\gamma}}{\hbar}(\mathbf{v^{\chi}_k}\cdot\mathbf{\Omega}^{\chi}_k)\right]\nonumber\\
 &= \sum_{\chi' \mathbf{k}'}{\mathbf{W}^{\chi \chi'}_{\mathbf{k k'}}}{(\Lambda^{1, \chi'}_{\mathbf{k'}}-\Lambda^{1, \chi}_{\mathbf{k}})},
\label{boltzman_in_terms_lambda}  
\end{align}
\begin{align}
&\frac{\partial}{\partial \epsilon^\chi_\mathbf{k}} \left(\frac{\partial f_0^\chi}{\partial \epsilon^\chi_\mathbf{k}} ~\Lambda^{1, \chi}_{\mathbf{k}}\right)
\left[{v^{\chi,z}_{\mathbf{k}}}+\frac{eB\sin{\gamma}}{\hbar}(\mathbf{v^{\chi}_k}\cdot\mathbf{\Omega}^{\chi}_k)\right]\nonumber\\
&= \frac{1}{\mathcal{D}^\chi}\sum\limits_{\chi'}\sum\limits_{\mathbf{k}'} W^{\chi \chi'}_{\mathbf{k}\mathbf{k}'} \times \nonumber\\
&\left[ \frac{\partial}{\partial \epsilon^{\chi'}_{\mathbf{k'}}} \left(\frac{\partial f_{0}^{\chi'}}{\partial \epsilon^{\chi'}{_\mathbf{k'}}} ~\Lambda^{1, \chi'}_{\mathbf{k'}}\right)~\Lambda^{2,\chi'}_{\mathbf{k}'}
- \frac{\partial}{\partial \epsilon^\chi_{\mathbf{k}}} \left(\frac{\partial f_{0}^\chi}{\partial \epsilon^\chi_{\mathbf{k}}} ~\Lambda^{1, \chi}_{\mathbf{k}}\right)~\Lambda^{2,\chi}_{\mathbf{k}} \right].
 \label{boltzman_in_terms_lambda_2}
\end{align}

\begin{table}[b]
\begin{ruledtabular}
\begin{tabular}{cccccccc}
 i &$\sigma_{i}$ &$\alpha_{i}$ &$\beta_{i}$  &&$\xi_{i}$\\
 \hline
 0 &$\mathbb{I}_{2\times2}$&+1 &+1 && +1\\
 1 &$\sigma_{x}$&-1 &+1 && -1\\
 2 &$\sigma_{y}$&+1 &-1 && -1\\
 3 &$\sigma_{z}$&-1 &-1 && +1\\
\end{tabular}
\caption{\label{tab:Table1}
The signs of $\alpha, \beta$, and $\xi$ are used in the expression of overlap of the Bloch wave function (Please see Eq.~\ref{Tau_invers}). $\sigma_{x, y, z}$ are the components of the Pauli spin vector, and $\mathbb{I}_{2\times2}$ is the identity matrix. It should be read as for $i=0$, $\sigma_{i}=\mathbb{I}_{2\times2}, \alpha_{i} = +1,\beta_{i} = +1$ and $\xi_{i} = +1$. We should follow a similar trend for $i=1, 2, 3$.}
\end{ruledtabular}
\end{table}
To get the distribution function up to the second order in $E$, we have to solve Eq.~\ref{boltzman_in_terms_lambda} and Eq.~\ref{boltzman_in_terms_lambda_2} for unknown function $\Lambda$. First we solve Eq.~\ref{boltzman_in_terms_lambda} for $\Lambda^{1,\chi}_{\mathbf{k}}$ and we later will return back to Eq.~\ref{boltzman_in_terms_lambda_2}. Here, we first define the band scattering rate as follows:
\begin{align}
\frac{1}{\tau^{\chi}_{\mathbf{k}}(\theta,\phi)}=\sum_{\chi'}\mathcal{V}\int\frac{d^3\mathbf{k'}}{(2\pi)^3}(\mathcal{D}^{\chi'}_{\mathbf{k}'})^{-1}\mathbf{W}^{\chi \chi'}_{\mathbf{k k'}}.
\label{Tau_invers}
\end{align}
$\mathbf{W}^{\chi \chi'}_{\mathbf{k k'}}$ is defined in Eq.~\ref{Fermi_gilden_rule} and the corresponding overlap of the Bloch wave-function will be given by the following expression,
$\mathcal{G}_{i}^{\chi\chi'}(\theta,\phi) = [1+\chi\chi'\xi_{i}(\cos{\theta}\cos{\theta'} + \alpha_{i}\sin{\theta}\sin{\theta'}\cos{\phi}\cos{\phi'} + \beta_{i}\sin{\theta}\sin{\theta'}\sin{\phi}\sin{\phi'}]$ with $i=0,1,2,3$ (Please see Tab.~\ref{tab:Table1}). Since the formalism remains identical for all types of impurity sites, we omit the corresponding index ($i$) for simplicity. Taking Berry phase into account and corresponding change in density of states, $\sum_{k}\longrightarrow \mathcal{V}\int\frac{d^3\mathbf{k}}{(2\pi)^3}\mathcal{D}^\chi(k)$, Eq.~\ref{boltzman_in_terms_lambda} becomes:
\begin{multline}
h^{\chi}_{\mu}(\theta,\phi) + \frac{\Lambda^{1,\chi}_{\mu}(\theta,\phi)}{\tau^{\chi}_{\mu}(\theta,\phi)}\\=\sum_{\chi'}\mathcal{V}\int\frac{d^3\mathbf{k}'}{(2\pi)^3} \mathcal{D}^{\chi'}(k') \mathcal{G}^{\chi\chi'} \mathbf{W}^{\chi \chi'}_{\mathbf{k k'}}\Lambda^{1,\chi'}_{\mu}(\theta',\phi').
\label{MB_in_term_Wkk'}
\end{multline}
Here, $h^{\chi}_{\mu}(\theta,\phi)=\mathcal{D}^{\chi}_{\mathbf{k}}[v^{\chi}_{z,\mathbf{k}}+\frac{eB\sin{\gamma}}{\hbar}(\mathbf{\Omega}^{\chi}_{k}\cdot \mathbf{v}^{\chi}_{\mathbf{k}})]$ and is evaluated at the Fermi energy. In low temperature limit, for a constant Fermi energy surface, Eq.~\ref{Tau_invers} and RHS of Eq.~\ref{MB_in_term_Wkk'} are reduced to integration over $\theta'$ and $\phi'$,
\begin{align}
\frac{1}{\tau^{\chi}_{\mu}(\theta,\phi)} =  \mathcal{V}\sum_{\chi'} \Pi^{\chi\chi'}\iint\frac{(k')^3\sin{\theta'}}{|\mathbf{v}^{\chi'}_{k'}\cdot{\mathbf{k'}^{\chi'}}|}d\theta'd\phi' ~\mathcal{G}^{\chi\chi'}(D^{\chi'}_{\mathbf{k'}})^{-1}.
\label{Tau_inv_int_thet_phi}
\end{align}
\begin{multline}
\mathcal{V}\sum_{\chi'} \Pi^{\chi\chi'}\iint f^{\chi'}(\theta',\phi') ~\mathcal{G}^{\chi \chi'} ~\Lambda^{1,\chi'}_{\mu}/\tau^{\chi'}_{\mu}~d\theta' d\phi'. 
\end{multline}
Where, $\Pi^{\chi \chi'} = N|U^{\chi\chi'}|^2 / 4\pi^2 \hbar^2$ and $f^{\chi} (\theta,\phi)=\frac{(k)^3}{|\mathbf{v}^\chi_{\mathbf{k}}\cdot \mathbf{k}^{\chi}|} \sin\theta (\mathcal{D}^\eta_{\mathbf{k}})^{-1} \tau^\chi_\mu(\theta,\phi)$. Using ansatz $\Lambda^{1,\chi}_{\mathbf{k}}=[d^{\chi}-h^{\chi}_{k'} + a^{\chi}\cos{\phi} +b^{\chi}\sin{\theta}\cos{\phi}+c^{\chi}\sin{\theta}\sin{\phi}]\tau^{\chi}_{\mu}(\theta,\phi)$ above equation can be written in following form:
\begin{multline}
d^{\chi}+a^{\chi}\cos{\phi}+b^{\chi}\sin{\theta}\cos{\phi}+c^{\chi} \sin{\theta}\sin{\phi}\\
=\sum_{\chi'}\mathcal{V}\Pi^{\chi\chi'}\iint f^{\chi'}(\theta',\phi') ~\mathcal{G}^{\chi\chi'} d\theta'd\phi'\\\times[d^{\chi'}-h^{\chi'}_{k'}+a^{\chi'}\cos{\theta'}+b^{\chi'}\sin{\theta'}\cos{\phi'}+c^{\chi'} \sin{\theta'}\sin{\phi'}].\\
\label{Boltzman_final}
\end{multline}
When aforementioned equation is explicitly put out, it appears as seven simultaneous equations that must be solved for eight variables. The particle number conservation provides an additional restriction.
\begin{align}
\sum\limits_{\chi}\sum\limits_{\mathbf{k}} g^{\chi}_\mathbf{k} = 0
\label{Eq_sumgk}
\end{align} 
Here, $g^{\chi}_\mathbf{k} = f_0+ g^{1,\chi}_\mathbf{k} + g^{2,\chi}_\mathbf{k}$. For eight unknowns ($d^{\pm 1}, a^{\pm 1}, b^{\pm 1}, c^{\pm 1}$), equations \ref{Boltzman_final} and \ref{Eq_sumgk} are simultaneously solved with Eq.~\ref{Tau_inv_int_thet_phi}. Due to intricate structure of equations, all two-dimensional integrals with respect to $\theta'$ and $\phi'$, simultaneous equations' solutions are carried out numerically. Returning back to the Eq.~\ref{boltzman_in_terms_lambda_2}, expanding each term and using the properties of Dirac delta function and its derivative, we get the following equation:
\begin{multline}
\mathcal{A}^{\chi}_\mathbf{k} (\theta,\phi) - \mathcal{B}^{\chi}_\mathbf{k} (\theta,\phi) \\
+ \frac{\partial}{\partial \epsilon^{\chi}_\mathbf{k}} \left( \frac{\Lambda^{1, \chi}_{\mathbf{k}}}{\tau^\chi_\mathbf{k} (\theta,\phi)} \Lambda^{2, \chi}_{\mathbf{k}} \right) - \frac{1}{\tau^\chi_\mathbf{k} (\theta,\phi)} \frac{\partial \Lambda^{1, \chi}_{\mathbf{k}}}{\partial \epsilon^{\chi}_\mathbf{k}} \Lambda^{2, \chi}_{\mathbf{k}} \\
= \sum_{\chi'} \left[ \frac{\partial}{\partial \epsilon^{\chi}_\mathbf{k}} \left( \int \frac{d^2 \mathbf{k'}}{(2\pi)^3} \mathcal{D}^{\chi'}_\mathbf{k'} W^{\chi \chi'}_{\mathbf{k} \mathbf{k'}} \Lambda^{1, \chi'}_{\mathbf{k'}} \Lambda^{2, \chi'}_{\mathbf{k'}} \right) \right. \\
\left. - \int \frac{d^2 \mathbf{k'}}{(2\pi)^3} \mathcal{D}^{\chi'}_\mathbf{k'} W^{\chi \chi'}_{\mathbf{k} \mathbf{k'}} \frac{\partial \Lambda^{1, \chi'}_{\mathbf{k'}}}{\partial \epsilon^{\chi'}_\mathbf{k'}} \Lambda^{2,\chi'}_{\mathbf{k'}} \right],
\label{eq:boltzman_in_terms_lambda_2_2}
\end{multline}
where, $\mathcal{A}^\chi_\mathbf{k} (\theta,\phi)= \frac{\partial }{\partial \epsilon^\chi_\mathbf{k}}\left[\mathcal{D}^{\chi}_{\mathbf{k}}(v^{\chi}_{z,\mathbf{k}}+\frac{eB\sin{\gamma}}{\hbar}(\mathbf{\Omega}^{\chi}_{k}\cdot \mathbf{v}^{\chi}_{\mathbf{k}}))\Lambda^{1,\chi}_{\mathbf{k}}\right]$ and $\mathcal{B}^\chi_\mathbf{k} (\theta,\phi) = \mathcal{D}^{\chi}_{\mathbf{k}}[v^{\chi}_{z,\mathbf{k}}+\frac{eB\sin{\gamma}}{\hbar}(\mathbf{\Omega}^{\chi}_{k}\cdot \mathbf{v}^{\chi}_{\mathbf{k}})] \left(\frac{\partial \Lambda^{1, \chi}_{\mathbf{k}}}{\partial \epsilon^\chi_\mathbf{k}}\right)$. $\tau^\chi_\mathbf{k} (\theta,\phi)$ is the same as defined in the Eq.~\ref{Tau_invers} and this Eq.~\ref{eq:boltzman_in_terms_lambda_2_2} has to be solved for $\Lambda^{2, \chi}_{\mathbf{k}}$ using $\Lambda^{1, \chi}_{\mathbf{k}}$ at Fermi level.\\

Eq.~\ref{boltzman_in_terms_lambda} can be solved for $\Lambda^{1,\chi}$, which can then be used to solve for $\Lambda^{2,\chi}$ in Eq.~\ref{eq:boltzman_in_terms_lambda_2_2}, and then the distribution function is evaluated using Eq.~\ref{Eq:g1} and ~\ref{Eq:g2} along $f^{\chi}_\mathbf{k} = f_{0} + g^{1, \chi}_{\mathbf{k}} + g^{2, \chi}_{\mathbf{k}} + O(E^3)$. Once the distribution function is evaluated, the current density can be evaluated as:
\begin{align}
    \mathbf{J}=-e\sum_{\chi,\mathbf{k}} f^{\chi}_{\mathbf{k}} \dot{\mathbf{r}}^{\chi}.
    \label{Eq:J_formula}
\end{align}
We primarily focus on the second-order anomalous Hall response induced by the chiral anomaly, which is given by, 
\begin{align}
    \mathbf{J}^\mathrm{CNLH}=-\frac{e^2}{\hbar}\sum_{\chi,\mathbf{k}} \mathcal{D}^\chi_\mathbf{k} g^{1,\chi}_{\mathbf{k}}  (\mathbf{E}\times \mathbf{\Omega}^\chi_\mathbf{k})
    \label{Eq:CNLH_formula}
\end{align}
To evaluate all the different responses, $\mathbf{J}^{\mathrm{CNLH}}$ is written as \cite{yao2024geometrical}:
\begin{align}
    J^{\mathrm{CNLH}}_{\alpha} = \sum_{\chi=\pm 1}\sum_{\beta \gamma}\sigma^{\chi}_{\alpha\beta \gamma} E_{\beta}E_{\gamma},
    \label{Eq:CNLH_componetns_formula}
\end{align}
with, $\alpha, \beta, \gamma = \{x,y,z\}$. A comparison between Eq.~\ref{Eq:CNLH_formula} and Eq.~\ref{Eq:CNLH_componetns_formula} allows us to identify the individual components of the third-rank nonlinear conductivity tensor, $\sigma^{\chi}_{\alpha\beta\gamma}$. For an applied electric field $\mathbf{E} = E \hat{z}$, the anomalous velocity, given by $\mathbf{v}^{\chi}_{\mathrm{anom}} \sim \mathbf{E} \times \mathbf{\Omega}^{\chi}_{\mathbf{k}}$, lies within the $xy$-plane. As the magnetic field is rotated within the $xz$-plane, the measurable Hall response manifests along the $y$-direction, corresponding to the $\sigma_{yzz}$ component. 
The contribution of $\mathcal{O}(E^2)$ to the chiral anomaly-induced nonlinear Hall current $\mathbf{J}^{\mathrm{CNLH}}$ arises through the following relations: $g^{1, \chi}_\mathbf{k} \sim \mathbf{E} \cdot \mathbf{\Lambda}^{1, \chi}_{\mathbf{k}}$, $g^{2, \chi}_\mathbf{k} \sim \frac{\partial}{\partial \epsilon^\chi_\mathbf{k}} \left(\frac{\partial f_0^\chi}{\partial \epsilon^\chi_\mathbf{k}} \, \mathbf{\Lambda}^{1, \chi}_{\mathbf{k}} \cdot \mathbf{E} \right) \, \mathbf{\Lambda}^{2,\chi}_{\mathbf{k}} \cdot \mathbf{E}$, and the anomalous velocity $\mathbf{v}^{\chi}_a \sim \mathbf{E} \times \mathbf{\Omega}^{\chi}_{\mathbf{k}}$. The vector $\mathbf{\Lambda}^{1, \chi}_{\mathbf{k}}$ can generally be expressed as $\mathbf{\Lambda}^{1, \chi}_{\mathbf{k}} = \varphi(\theta, \phi, k(B,\gamma))$, where $\varphi$ denotes an arbitrary function determined by the specific form of $h^{\chi}_{\mathbf{k}}$. Upon examining the product of $\mathbf{v}^{\chi}_a$ and the distribution function $g^{1,\chi}_\mathbf{k}$ in Eq.~\ref{Eq:CNLH_formula}, it becomes evident that the leading $\mathcal{O}(E^2)$ contribution to $\mathbf{J}^{\mathrm{CNLH}}$ stems from $g^{1, \chi}_\mathbf{k}$ in the direction perpendicular to the electric field $\mathbf{E}$. Another Hall contribution $\sim E^2$ is also possible due to the multiplication of the velocity term $\mathbf{v_k^\chi}$  and second-order distribution function in $E$ (Eq.~\ref{Eq:J_formula}). However, it requires solving Eq.~\ref{boltzman_in_terms_lambda_2} to determine $\mathbf{\Lambda}^{2, \chi}_{\mathbf{k}}$, which we leave as an open problem for future investigation.
It is important to note that the magnetic field is oriented in the $xz$-plane, resulting in a nonlinear Hall response along the $y$-direction, i.e., $\sigma_{yzz}$. While an additional non-linear Hall component—namely, the planar Hall conductivity $\sigma_{xzz}$ and a longitudinal component $\sigma_{zzz}$ (both of which require $\mathbf{\Lambda}^{2, \chi}_{\mathbf{k}}$) is also possible, it is not considered in the current study and is left for future work.

\section{Changes in Maxwell-Boltzmann transport (MBT) theory due to shift of origin}
\label{App_MBT theory due to shift of origin}
\begin{figure}
    \centering
    \includegraphics[width=.90\columnwidth]
    {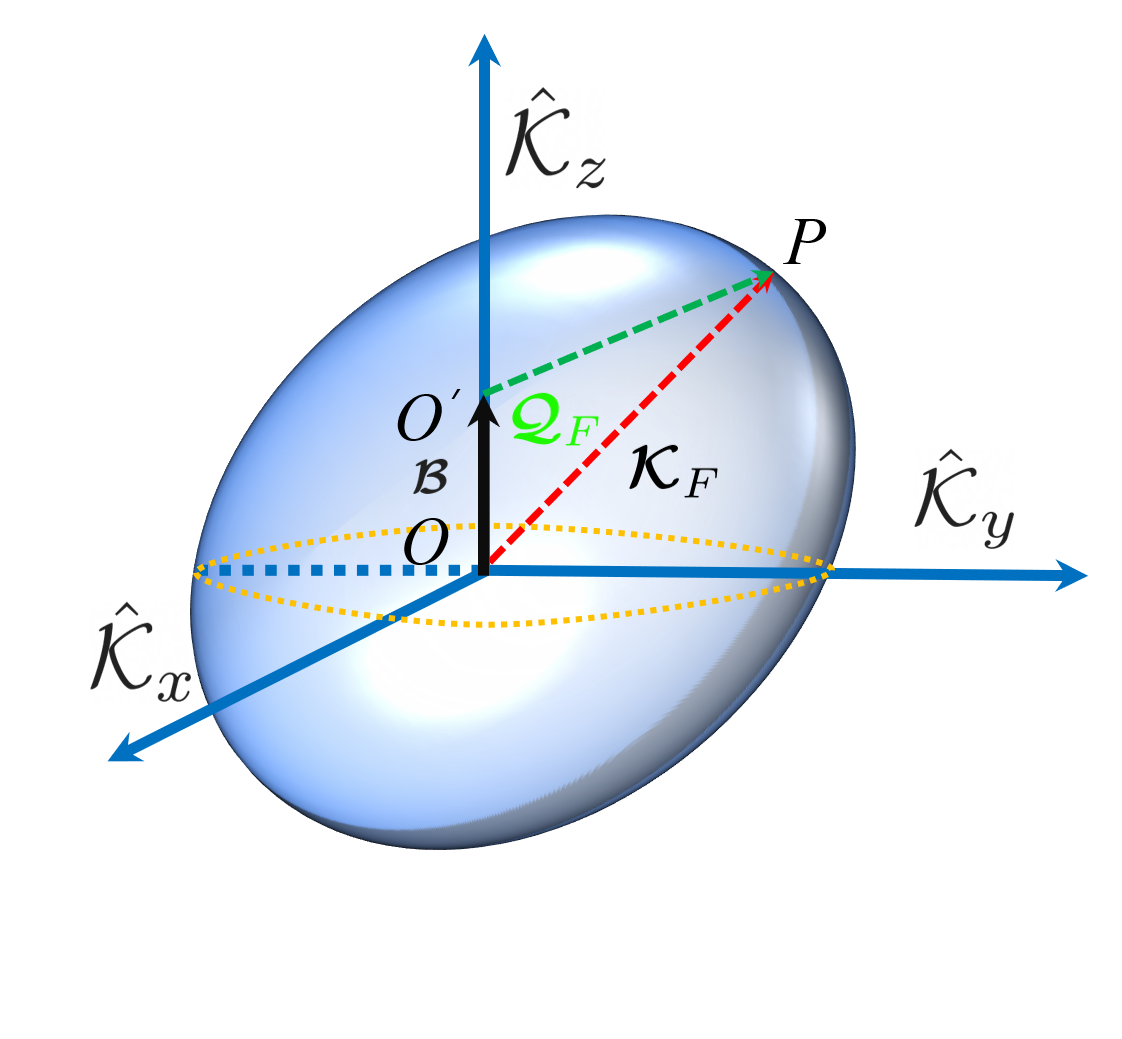}
    \caption{Schematic representation of a Fermi surface $\mathcal{E}_{F}$ in $\boldsymbol{\mathcal{K}}$-space. $O$ is old origin and $O^\prime$ is new origin shifted by $\boldsymbol{\mathcal{B}}$. $\boldsymbol{\mathcal{K}}_{F}$ and $\boldsymbol{\mathcal{Q}}_{F}$ are Fermi wave vector of a point $P$ on the Fermi surface with respect to $O$ and $O^\prime$ respectively. So, $\boldsymbol{\mathcal{K}}_{F} = \boldsymbol{\mathcal{B}} + \boldsymbol{\mathcal{Q}}_{F}$.}
\label{fig:Shift_of_origine}
\end{figure}

The modifications to the Maxwell-Boltzmann transport (MBT) formalism manifest explicitly in Eq.~\ref{Collision_integral} as a consequence of a shift in the momentum-space origin. This transformation affects the limits of integration over the constant-energy surface in momentum space. To illustrate this, consider Fig.~\ref{fig:Shift_of_origine}, where a constant Fermi surface at energy $\mathcal{E}_F$ is depicted in momentum space. The position vector of a representative point $P$ on the Fermi surface, measured from the original and shifted origins $O$ and $O'$, respectively, are related via: $\boldsymbol{\mathcal{K}}_{\mathrm{F}} = \boldsymbol{\mathcal{B}} + \boldsymbol{\mathcal{Q}}_{\mathrm{F}},$
where $\boldsymbol{\mathcal{B}}$ encodes both the magnitude and direction of the origin shift in momentum space. This shift alters the parametrization of the constant energy surface, and the momentum-space integral must be evaluated accordingly. The energy integral is computed using the following transformation relation:

\begin{align}
    \oint_{\mathcal{D}_{\boldsymbol{\mathcal{K}}}} \mathcal{F}_1(\mathbf{\boldsymbol{\mathcal{K}}}) \,d^3\mathbf{\boldsymbol{\mathcal{K}}} = \oint_{\mathcal{D}_{\boldsymbol{\mathcal{Q}}}} \mathcal{F}_2(\mathbf{\boldsymbol{\mathcal{Q}}}) \,d^3\mathbf{\boldsymbol{\mathcal{Q}}},
\end{align}
where, $\mathcal{F}_1(\mathbf{\boldsymbol{\mathcal{K}}})$ and $\mathcal{F}_2(\mathbf{\boldsymbol{\mathcal{Q}}})$ generic functions of $\mathbf{\boldsymbol{\mathcal{K}}}$ and $\mathbf{\boldsymbol{\mathcal{Q}}}$. The domains of integration are defined over iso-energetic surfaces in momentum space as: $\mathcal{D}_{\boldsymbol{\mathcal{K}}} = \left\{ \mathcal{K}(\theta, \phi) ~\big|~ \mathcal{K}(\theta, \phi) = \mathcal{K}_\mathrm{F},~ \mathcal{E}(\mathcal{K}(\theta, \phi)) = \mathcal{E}_\mathrm{F} \right\}, \quad 
\mathcal{Q}_{\boldsymbol{\mathcal{Q}}} = \left\{ \mathcal{Q}(\theta, \phi) ~\big|~ \mathcal{Q}(\theta, \phi) = \mathcal{Q}_\mathrm{F},~ \mathcal{E}(\mathcal{Q}(\theta, \phi)) = \mathcal{E}_\mathrm{F} \right\}$. These correspond to constant energy contours defined on the Fermi surface. The dispersion relation derived from Eq.~\ref{eq:H_SOC_3_in_q} yields isoenergy manifolds that are analogous to those reported in Ref.~\cite{varma2024magnetotransport} for spin-orbit coupled noncentrosymmetric Weyl metals (SOC-NCMs), thereby enabling a direct comparison of their topological and transport features. 

\label{SUbsection_general_form_BC}
\section{Berry curvature for a general two-band Hamiltonian.}
\label{App_Berry curvature for a 2 band}
Let's take general form of two-band Hamiltonian,
\begin{align}
    H_{2\times2} (\mathbf{k}) = \boldsymbol{\mathcal{W}} (\boldsymbol{k}) \cdot \boldsymbol{\sigma}.
    \label{Eq: General_two_band}
\end{align}
Here, $\mathbf{k}$ is wave vector and $\boldsymbol{\sigma}$ is vector of Pauli matrices. $\boldsymbol{\mathcal{W}} (\mathbf{k}) = \mathcal{W} ~(\sin{\alpha} \cos{\beta}, ~\sin{\alpha} \sin{\beta}, ~\cos{\alpha})$ with $0\leq \alpha \leq \pi$, $0 \leq \beta \leq 2\pi$. Then, the Eq.~\ref{Eq: General_two_band} can be written as,
\begin{align}
H_{2\times2} (\mathbf{k}) = \mathcal{W}~
\begin{bmatrix}
\cos{\alpha} & \sin{\alpha} ~e^{-i \beta}\\
\sin{\alpha} ~e^{i \beta} & -\cos{\alpha}
\end{bmatrix} 
\end{align}
This can be diagonalized easily to get eigenvalues ($\pm |\mathcal{W}(\mathbf{k})|$) and eigenfunctions ($\ket{u^{+}}^{T}=[e^{-i\beta} ~\cos(\alpha /2),  ~\sin(\beta/2)], ~\ket{u^{-}}^{T}=[-e^{-i\beta} ~\sin(\alpha /2), ~\cos(\beta/2)]$). We used the following formula to calculate the Berry curvature (BC),
\begin{align}
    \Omega_{xy} = -2~ \mathrm{Im} \biggr\langle \frac{\partial u}{\partial k_{x}} \biggr | \frac{\partial u}{\partial k_{x}} \biggr \rangle.
\end{align}
Using chain rule and relation, $\cos{\alpha} = \mathcal{W}_{z}/\mathcal{W}$ and $\tan{\beta} =  \mathcal{W}_{y}/\mathcal{W}_{x}$ (where $\mathcal{W}_x, \mathcal{W}_y, \mathcal{W}_z$ are components of $\boldsymbol{\mathcal{W}}$  and $\mathcal{W}=|\boldsymbol{\mathcal{W}}|$), one may obtain the general form of the BC as;
\begin{align}
    \Omega_{ij} = \frac{\sin{\alpha}}{2} \left ( \frac{\partial \alpha}{\partial k_{i}}  \frac{\partial \beta}{\partial k_{j}} - \frac{\partial \alpha}{\partial k_{j}}  \frac{\partial \beta}{\partial k_{i}}\right),
    \label{eq: BC_genaral_form}
\end{align}
with, $i,j=x,y,z$. A similar calculation is also possible for the OMM in terms of $\alpha$ and $\beta$.

\bibliography{biblio.bib}

@article{chang2018topological,
  title={Topological quantum properties of chiral crystals},
  author={Chang, Guoqing and Wieder, Benjamin J and Schindler, Frank and Sanchez, Daniel S and Belopolski, Ilya and Huang, Shin-Ming and Singh, Bahadur and Wu, Di and Chang, Tay-Rong and Neupert, Titus and others},
  journal={Nature materials},
  volume={17},
  number={11},
  pages={978--985},
  year={2018},
  publisher={Nature Publishing Group UK London}
}

@article{hasan2021weyl,
  title={Weyl, Dirac and high-fold chiral fermions in topological quantum matter},
  author={Hasan, M Zahid and Chang, Guoqing and Belopolski, Ilya and Bian, Guang and Xu, Su-Yang and Yin, Jia-Xin},
  journal={Nature Reviews Materials},
  volume={6},
  number={9},
  pages={784--803},
  year={2021},
  publisher={Nature Publishing Group UK London}
}

@misc{sarkar2025symmetrydrivenintrinsicnonlinearpure,
      title={Symmetry-driven Intrinsic Nonlinear Pure Spin Hall Effect}, 
      author={Sayan Sarkar and Sunit Das and Amit Agarwal},
      year={2025},
      eprint={2502.18226},
      archivePrefix={arXiv},
      primaryClass={cond-mat.mes-hall},
      url={https://arxiv.org/abs/2502.18226}, 
}

@article{bradlyn2016beyond,
  title={Beyond Dirac and Weyl fermions: Unconventional quasiparticles in conventional crystals},
  author={Bradlyn, Barry and Cano, Jennifer and Wang, Zhijun and Vergniory, MG and Felser, C and Cava, Robert Joseph and Bernevig, B Andrei},
  journal={Science},
  volume={353},
  number={6299},
  pages={aaf5037},
  year={2016},
  publisher={American Association for the Advancement of Science}
}

@article{ahmad2023longitudinal,
  title={Longitudinal magnetoconductance and the planar Hall conductance in inhomogeneous Weyl semimetals},
  author={Ahmad, Azaz and Raman, Karthik V and Tewari, Sumanta and Sharma, G},
  journal={Physical Review B},
  volume={107},
  number={14},
  pages={144206},
  year={2023},
  publisher={APS}
}

@article{das2023chiral,
  title={Chiral anomalies in three-dimensional spin-orbit coupled metals: Electrical, thermal, and gravitational anomalies},
  author={Das, Sunit and Das, Kamal and Agarwal, Amit},
  journal={Physical Review B},
  volume={108},
  number={4},
  pages={045405},
  year={2023},
  publisher={APS}
}

@article{ahmad2021longitudinal,
  title={Longitudinal magnetoconductance and the planar Hall effect in a lattice model of tilted Weyl fermions},
  author={Ahmad, Azaz and Sharma, Gargee},
  journal={Physical Review B},
  volume={103},
  number={11},
  pages={115146},
  year={2021},
  publisher={APS}
}

@article{armitage2018weyl,
  title={Weyl and Dirac semimetals in three-dimensional solids},
  author={Armitage, NP and Mele, EJ and Vishwanath, Ashvin},
  journal={Reviews of Modern Physics},
  volume={90},
  number={1},
  pages={015001},
  year={2018},
  publisher={APS}
}

@article{sharma2017nernst,
  title={Nernst effect in Dirac and inversion-asymmetric Weyl semimetals},
  author={Sharma, Gargee and Moore, Christopher and Saha, Subhodip and Tewari, Sumanta},
  journal={Physical Review B},
  volume={96},
  number={19},
  pages={195119},
  year={2017},
  publisher={APS}
}

@article{sharma2019transverse,
  title={Transverse thermopower in Dirac and Weyl semimetals},
  author={Sharma, Gargee and Tewari, Sumanta},
  journal={Physical Review B},
  volume={100},
  number={19},
  pages={195113},
  year={2019},
  publisher={APS}
}

@article{liang2017anomalous,
  title={Anomalous Nernst effect in the dirac semimetal Cd 3 As 2},
  author={Liang, Tian and Lin, Jingjing and Gibson, Quinn and Gao, Tong and Hirschberger, Max and Liu, Minhao and Cava, Robert Joseph and Ong, Nai Phuan},
  journal={Physical Review Letters},
  volume={118},
  number={13},
  pages={136601},
  year={2017},
  publisher={APS}
}

@article{burkov2011weyl,
  title={Weyl semimetal in a topological insulator multilayer},
  author={Burkov, AA and Balents, Leon},
  journal={Physical Review Letters},
  volume={107},
  number={12},
  pages={127205},
  year={2011},
  publisher={APS}
}

@article{sharma2023decoupling,
  title={Decoupling intranode and internode scattering in Weyl fermions},
  author={Sharma, Gargee and Nandy, Snehasish and Raman, Karthik V and Tewari, Sumanta},
  journal={Physical Review B},
  volume={107},
  number={11},
  pages={115161},
  year={2023},
  publisher={APS}
}

@article{zyuzin2012topological,
  title={Topological response in Weyl semimetals and the chiral anomaly},
  author={Zyuzin, AA and Burkov, AA},
  journal={Physical Review B},
  volume={86},
  number={11},
  pages={115133},
  year={2012},
  publisher={APS}
}

@article{wan2011topological,
  title={Topological semimetal and Fermi-arc surface states in the electronic structure of pyrochlore iridates},
  author={Wan, Xiangang and Turner, Ari M and Vishwanath, Ashvin and Savrasov, Sergey Y},
  journal={Physical Review B},
  volume={83},
  number={20},
  pages={205101},
  year={2011},
  publisher={APS}
}

@article{yang2011quantum,
  title={Quantum Hall effects in a Weyl semimetal: Possible application in pyrochlore iridates},
  author={Yang, Kai-Yu and Lu, Yuan-Ming and Ran, Ying},
  journal={Physical Review B},
  volume={84},
  number={7},
  pages={075129},
  year={2011},
  publisher={APS}
}

@article{xiao2010berry,
  title={Berry phase effects on electronic properties},
  author={Xiao, Di and Chang, Ming-Che and Niu, Qian},
  journal={Reviews of modern physics},
  volume={82},
  number={3},
  pages={1959},
  year={2010},
  publisher={APS}
}

@techreport{nielsen1981no,
  title={No-go theorum for regularizing chiral fermions},
  author={Nielsen, Holger Bech and Ninomiya, Masao},
  year={1981},
  institution={Science Research Council}
}

@article{fukushima2008chiral,
  title={Chiral magnetic effect},
  author={Fukushima, Kenji and Kharzeev, Dmitri E and Warringa, Harmen J},
  journal={Physical Review D},
  volume={78},
  number={7},
  pages={074033},
  year={2008},
  publisher={APS}
}

@article{nielsen1983adler,
  title={The Adler-Bell-Jackiw anomaly and Weyl fermions in a crystal},
  author={Nielsen, Holger Bech and Ninomiya, Masao},
  journal={Physics Letters B},
  volume={130},
  number={6},
  pages={389--396},
  year={1983},
  publisher={Elsevier}
}

@article{burkov2014anomalous,
  title={Anomalous Hall effect in Weyl metals},
  author={Burkov, AA},
  journal={Physical Review Letters},
  volume={113},
  number={18},
  pages={187202},
  year={2014},
  publisher={APS}
}

@article{goswami2015optical,
  title={Optical activity as a test for dynamic chiral magnetic effect of weyl semimetals},
  author={Goswami, Pallab and Sharma, Gargee and Tewari, Sumanta},
  journal={Physical Review B},
  volume={92},
  number={16},
  pages={161110},
  year={2015},
  publisher={APS}
}

@article{goswami2013axionic,
  title={Axionic field theory of (3+ 1)-dimensional Weyl semimetals},
  author={Goswami, Pallab and Tewari, Sumanta},
  journal={Physical Review B},
  volume={88},
  number={24},
  pages={245107},
  year={2013},
  publisher={APS}
}

@article{bell1969pcac,
  title={A PCAC puzzle: $\pi$ 0→ $\gamma$$\gamma$ in the $\sigma$-model},
  author={Bell, John S and Jackiw, Roman},
  journal={Il Nuovo Cimento A (1965-1970)},
  volume={60},
  number={1},
  pages={47--61},
  year={1969},
  publisher={Springer}
}

@article{aji2012adler,
  title={Adler-Bell-Jackiw anomaly in Weyl semimetals: Application to pyrochlore iridates},
  author={Aji, Vivek},
  journal={Physical Review B},
  volume={85},
  number={24},
  pages={241101},
  year={2012},
  publisher={APS}
}

@article{adler1969axial,
  title={Axial-vector vertex in spinor electrodynamics},
  author={Adler, Stephen L},
  journal={Physical Review},
  volume={177},
  number={5},
  pages={2426},
  year={1969},
  publisher={APS}
}

@article{zyuzin2012weyl,
  title={Weyl semimetal with broken time reversal and inversion symmetries},
  author={Zyuzin, AA and Wu, Si and Burkov, AA},
  journal={Physical Review B},
  volume={85},
  number={16},
  pages={165110},
  year={2012},
  publisher={APS}
}

@article{son2012berry,
  title={Berry curvature, triangle anomalies, and the chiral magnetic effect in Fermi liquids},
  author={Son, Dam Thanh and Yamamoto, Naoki},
  journal={Physical Review Letters},
  volume={109},
  number={18},
  pages={181602},
  year={2012},
  publisher={APS}
}

@article{son2013chiral,
  title={Chiral anomaly and classical negative magnetoresistance of Weyl metals},
  author={Son, DT and Spivak, BZ},
  journal={Physical Review B},
  volume={88},
  number={10},
  pages={104412},
  year={2013},
  publisher={APS}
}

@article{kim2014boltzmann,
  title={Boltzmann equation approach to anomalous transport in a Weyl metal},
  author={Kim, Ki-Seok and Kim, Heon-Jung and Sasaki, M},
  journal={Physical Review B},
  volume={89},
  number={19},
  pages={195137},
  year={2014},
  publisher={APS}
}

@article{sharma2016nernst,
  title={Nernst and magnetothermal conductivity in a lattice model of Weyl fermions},
  author={Sharma, Gargee and Goswami, Pallab and Tewari, Sumanta},
  journal={Physical Review B},
  volume={93},
  number={3},
  pages={035116},
  year={2016},
  publisher={APS}
}

@article{nandy2017chiral,
  title={Chiral anomaly as the origin of the planar Hall effect in Weyl semimetals},
  author={Nandy, S and Sharma, Gargee and Taraphder, A and Tewari, Sumanta},
  journal={Physical Review Letters},
  volume={119},
  number={17},
  pages={176804},
  year={2017},
  publisher={APS}
}

@article{knoll2020negative,
  title={Negative longitudinal magnetoconductance at weak fields in Weyl semimetals},
  author={Knoll, Andy and Timm, Carsten and Meng, Tobias},
  journal={Physical Review B},
  volume={101},
  number={20},
  pages={201402},
  year={2020},
  publisher={APS}
}

@article{zhang2016linear,
  title={Linear magnetoconductivity in an intrinsic topological Weyl semimetal},
  author={Zhang, Song-Bo and Lu, Hai-Zhou and Shen, Shun-Qing},
  journal={New Journal of Physics},
  volume={18},
  number={5},
  pages={053039},
  year={2016},
  publisher={IOP Publishing}
}

@article{sharma2017chiral,
  title={Chiral anomaly and longitudinal magnetotransport in type-II Weyl semimetals},
  author={Sharma, Gargee and Goswami, Pallab and Tewari, Sumanta},
  journal={Physical Review B},
  volume={96},
  number={4},
  pages={045112},
  year={2017},
  publisher={APS}
}

@article{sharma2020sign,
  title={Sign of longitudinal magnetoconductivity and the planar Hall effect in Weyl semimetals},
  author={Sharma, Gargee and Nandy, S and Tewari, Sumanta},
  journal={Physical Review B},
volume={102},
number={20},
pages={205107},
year={2020},
publisher={APS}
}

@article{Yan_2017,
   title={Topological Materials: Weyl Semimetals},
   volume={8},
   ISSN={1947-5462},
   url={http://dx.doi.org/10.1146/annurev-conmatphys-031016-025458},
   DOI={10.1146/annurev-conmatphys-031016-025458},
   number={1},
   journal={Annual Review of Condensed Matter Physics},
   publisher={Annual Reviews},
   author={Yan, Binghai and Felser, Claudia},
   year={2017},
   month=mar, pages={337–354} }

@book{abers2004quantum,
  title={Quantum Mechanics},
  author={Abers, E.S.},
  isbn={9780131461000},
  lccn={2003051437},
  url={https://books.google.co.in/books?id=_aBkQgAACAAJ},
  year={2004},
  publisher={Pearson Education}
}

@ARTICLE{1929ZPhy...56..330W,
       author = {{Weyl}, Hermann},
        title = "{Elektron und Gravitation. I}",
      journal = {Zeitschrift fur Physik},
         year = 1929,
        month = may,
       volume = {56},
       number = {5-6},
        pages = {330-352},
          doi = {10.1007/BF01339504},
       adsurl = {https://ui.adsabs.harvard.edu/abs/1929ZPhy...56..330W}
}

@article{PhysRevB.103.045105,
  title = {Nonlinear Hall effect in Weyl semimetals induced by chiral anomaly},
  author = {Li, Rui-Hao and Heinonen, Olle G. and Burkov, Anton A. and Zhang, Steven S.-L.},
  journal = {Phys. Rev. B},
  volume = {103},
  issue = {4},
  pages = {045105},
  numpages = {11},
  year = {2021},
  month = {Jan},
  publisher = {American Physical Society},
  doi = {10.1103/PhysRevB.103.045105},
  url = {https://link.aps.org/doi/10.1103/PhysRevB.103.045105}
}

@article{PhysRevB.104.205124,
  title = {Chiral anomaly induced nonlinear Hall effect in semimetals with multiple Weyl points},
  author = {Nandy, Snehasish and Zeng, Chuanchang and Tewari, Sumanta},
  journal = {Phys. Rev. B},
  volume = {104},
  issue = {20},
  pages = {205124},
  numpages = {7},
  year = {2021},
  month = {Nov},
  publisher = {American Physical Society},
  doi = {10.1103/PhysRevB.104.205124},
  url = {https://link.aps.org/doi/10.1103/PhysRevB.104.205124}
}

@article{PhysRevB.105.125131,
  title = {Chiral anomaly induced nonlinear Nernst and thermal Hall effects in Weyl semimetals},
  author = {Zeng, Chuanchang and Nandy, Snehasish and Tewari, Sumanta},
  journal = {Phys. Rev. B},
  volume = {105},
  issue = {12},
  pages = {125131},
  numpages = {11},
  year = {2022},
  month = {Mar},
  publisher = {American Physical Society},
  doi = {10.1103/PhysRevB.105.125131},
  url = {https://link.aps.org/doi/10.1103/PhysRevB.105.125131}
}

@article{PhysRevB.107.115161,
  title = {Decoupling intranode and internode scattering in Weyl fermions},
  author = {Sharma, G. and Nandy, Snehasish and Raman, Karthik V. and Tewari, Sumanta},
  journal = {Phys. Rev. B},
  volume = {107},
  issue = {11},
  pages = {115161},
  numpages = {6},
  year = {2023},
  month = {Mar},
  publisher = {American Physical Society},
  doi = {10.1103/PhysRevB.107.115161},
  url = {https://link.aps.org/doi/10.1103/PhysRevB.107.115161}
}

@article{PhysRev.95.1154,
  title = {Hall Effect in Ferromagnetics},
  author = {Karplus, Robert and Luttinger, J. M.},
  journal = {Phys. Rev.},
  volume = {95},
  issue = {5},
  pages = {1154--1160},
  numpages = {0},
  year = {1954},
  month = {Sep},
  publisher = {American Physical Society},
  doi = {10.1103/PhysRev.95.1154},
  url = {https://link.aps.org/doi/10.1103/PhysRev.95.1154}
}

@article{PhysRevB.97.201404,
  title = {Rotational strain in Weyl semimetals: A continuum approach},
  author = {Arjona, Vicente and Vozmediano, Mar\'{\i}a A. H.},
  journal = {Phys. Rev. B},
  volume = {97},
  issue = {20},
  pages = {201404},
  numpages = {4},
  year = {2018},
  month = {May},
  publisher = {American Physical Society},
  doi = {10.1103/PhysRevB.97.201404},
  url = {https://link.aps.org/doi/10.1103/PhysRevB.97.201404}
}

@article{
doi:10.1126/sciadv.1603266,
author = {Su-Yang Xu  and Nasser Alidoust  and Guoqing Chang  and Hong Lu  and Bahadur Singh  and Ilya Belopolski  and Daniel S. Sanchez  and Xiao Zhang  and Guang Bian  and Hao Zheng  and Marious-Adrian Husanu  and Yi Bian  and Shin-Ming Huang  and Chuang-Han Hsu  and Tay-Rong Chang  and Horng-Tay Jeng  and Arun Bansil  and Titus Neupert  and Vladimir N. Strocov  and Hsin Lin  and Shuang Jia  and M. Zahid Hasan },
title = {Discovery of Lorentz-violating type II Weyl fermions in LaAlGe},
journal = {Science Advances},
volume = {3},
number = {6},
pages = {e1603266},
year = {2017},
doi = {10.1126/sciadv.1603266},
URL = {https://www.science.org/doi/abs/10.1126/sciadv.1603266},
eprint = {https://www.science.org/doi/pdf/10.1126/sciadv.1603266}}

@article{li2021nonlinear,
  title={Nonlinear Hall effect in Weyl semimetals induced by chiral anomaly},
  author={Li, Rui-Hao and Heinonen, Olle G and Burkov, Anton A and Zhang, Steven S-L},
  journal={Physical Review B},
  volume={103},
  number={4},
  pages={045105},
  year={2021},
  publisher={APS}
}

@article{nandy2021chiral,
  title={Chiral anomaly induced nonlinear Hall effect in semimetals with multiple Weyl points},
  author={Nandy, Snehasish and Zeng, Chuanchang and Tewari, Sumanta},
  journal={Physical Review B},
  volume={104},
  number={20},
  pages={205124},
  year={2021},
  publisher={APS}
}

@article{zeng2022chiral,
  title={Chiral anomaly induced nonlinear Nernst and thermal Hall effects in Weyl semimetals},
  author={Zeng, Chuanchang and Nandy, Snehasish and Tewari, Sumanta},
  journal={Physical Review B},
  volume={105},
  number={12},
  pages={125131},
  year={2022},
  publisher={APS}
}

@article{yakunin2010spin,
  title={Spin splittings in the n-HgTe/CdxHg1- xTe (013) quantum well with inverted band structure},
  author={Yakunin, MV and Podgornykh, SM and Mikhailov, NN and Dvoretsky, SA},
  journal={Physica E: Low-dimensional Systems and Nanostructures},
  volume={42},
  number={4},
  pages={948--951},
  year={2010},
  publisher={Elsevier}
}

@article{xie2021kramers,
  title={Kramers nodal line metals},
  author={Xie, Ying-Ming and Gao, Xue-Jian and Xu, Xiao Yan and Zhang, Cheng-Ping and Hu, Jin-Xin and Gao, Jason Z and Law, Kam Tuen},
  journal={Nature communications},
  volume={12},
  number={1},
  pages={3064},
  year={2021},
  publisher={Nature Publishing Group UK London}
}

@article{yao2024geometrical,
  title={Geometrical nonlinear Hall effect induced by Lorentz force},
  author={Yao, Junjie and Liu, Yizhou and Duan, Wenhui},
  journal={Physical Review B},
  volume={110},
  number={11},
  pages={115123},
  year={2024},
  publisher={APS}
}

@article{verma2019thermoelectric,
  title={Thermoelectric and optical probes for a Fermi surface topology change in noncentrosymmetric metals},
  author={Verma, Sonu and Biswas, Tutul and Ghosh, Tarun Kanti},
  journal={Physical Review B},
  volume={100},
  number={4},
  pages={045201},
  year={2019},
  publisher={APS}
}

@article{cheon2022chiral,
  title={Chiral anomaly in noncentrosymmetric systems induced by spin-orbit coupling},
  author={Cheon, Suik and Cho, Gil Young and Kim, Ki-Seok and Lee, Hyun-Woo},
  journal={Physical Review B},
  volume={105},
  number={18},
  pages={L180303},
  year={2022},
  publisher={APS}
}

@article{gao2022chiral,
  title={Chiral anomaly in non-relativistic systems: berry curvature and chiral kinetic theory},
  author={Gao, Lan-Lan and Huang, Xu-Guang},
  journal={Chinese Physics Letters},
  volume={39},
  number={2},
  pages={021101},
  year={2022},
  publisher={IOP Publishing}
}

@article{varma2024magnetotransport,
  title={Magnetotransport in spin-orbit coupled noncentrosymmetric and Weyl metals},
  author={Varma, Gautham and Ahmad, Azaz and Tewari, Sumanta and Sharma, Gargee},
  journal={Physical Review B},
  volume={109},
  number={16},
  pages={165114},
  year={2024},
  publisher={APS}
}

@article{ahmad2024geometry,
  title={Geometry, anomaly, topology, and transport in Weyl fermions},
  author={Ahmad, Azaz and Varma, Gautham and Sharma, Gargee},
  journal={Journal of Physics: Condensed Matter},
  volume={37},
  number={4},
  pages={043001},
  year={2024},
  publisher={IOP Publishing}
}

@article{ahmad2025longitudinal,
  title={Longitudinal magnetoconductance of higher-pseudospin fermions},
  author={Ahmad, Azaz and Sharma, Gargee},
  journal={Physical Review B},
  volume={112},
  number={4},
  pages={045135},
  year={2025},
  publisher={APS}
}

@article{ahmad2025chiral,
  title={Chiral anomaly induced nonlinear Hall effect in three-dimensional chiral fermions},
  author={Ahmad, Azaz and K, Gautham Varma and Sharma, Gargee},
  journal={Physical Review B},
  volume={111},
  number={3},
  pages={035138},
  year={2025},
  publisher={APS}
}

@article{bevan1997momentum,
  title={Momentum creation by vortices in superfluid 3He as a model of primordial baryogenesis},
  author={Bevan, TDC and Manninen, AJ and Cook, JB and Hook, JR and Hall, HE and Vachaspati, Tanmay and Volovik, GE},
  journal={Nature},
  volume={386},
  number={6626},
  pages={689--692},
  year={1997},
  publisher={Nature Publishing Group UK London}
}

@article{mandal2025disentangling,
  title={Disentangling contributions to longitudinal magnetoconductivity for Kramers-Weyl nodes},
  author={Mandal, Ipsita},
  journal={arXiv preprint arXiv:2506.07913},
  year={2025}
}

@article{wieder2022topological,
  title={Topological materials discovery from crystal symmetry},
  author={Wieder, Benjamin J and Bradlyn, Barry and Cano, Jennifer and Wang, Zhijun and Vergniory, Maia G and Elcoro, Luis and Soluyanov, Alexey A and Felser, Claudia and Neupert, Titus and Regnault, Nicolas and others},
  journal={Nature Reviews Materials},
  volume={7},
  number={3},
  pages={196--216},
  year={2022},
  publisher={Nature Publishing Group UK London}
}

@article{zhang2024third,
  title={Third-order anomalous Hall response to the Fermi-surface topology in magnetic materials},
  author={Zhang, Xu and Sun, Kai and Meng, Zi Yang},
  journal={Physical Review B},
  volume={110},
  number={16},
  pages={165140},
  year={2024},
  publisher={APS}
}
\end{document}